\documentclass[prb,superscriptaddress,showpacs,preprint, amsmath, amssymb, aps , floatfix]{revtex4-1}

\usepackage{graphicx}
\usepackage{dcolumn}
\usepackage{bm}
\usepackage{appendix}

\newcommand{\overbar}[1]{\mkern 1.5mu\overline{\mkern-1.5mu#1\mkern-1.5mu}\mkern 1.5mu}

\begin{document}


\title{Low-Temperature Insulating Phase of the Si(111)--7$\times$7 Surface}

\author{S. Modesti}
 \affiliation{Dipartimento di Fisica, Universit\`a di Trieste, I-34127 Trieste, Italy}
 \affiliation{Istituto Officina dei Materiali, Consiglio Nazionale delle Ricerche, I-34149 Trieste, Italy}
\author{P. M. Sheverdyaeva} 
\affiliation{Istituto di Struttura della Materia, Consiglio Nazionale delle Ricerche, I-34149 Trieste, Italy}%
\author{P. Moras} 
\affiliation{Istituto di Struttura della Materia, Consiglio Nazionale delle Ricerche, I-34149 Trieste, Italy}%
\author{C. Carbone}
\affiliation{Istituto di Struttura della Materia, Consiglio Nazionale delle Ricerche, I-34149 Trieste, Italy}%
\author{M. Caputo}
\affiliation{Laboratoire de Physique des Solides, CNRS, Universit\'{e} Paris-Sud, Universit\'{e} Paris-Saclay, F-91405 Orsay Cedex, France}
\author{M. Marsi}
\affiliation{Laboratoire de Physique des Solides, CNRS, Universit\'{e} Paris-Sud, Universit\'{e} Paris-Saclay, F-91405 Orsay Cedex, France}
\author{E. Tosatti}
\affiliation{International School for Advanced Studies (SISSA), Via Bonomea 265, I-34136 Trieste, Italy}
\affiliation{CNR-IOM Democritos, Via Bonomea 265, I-34136 Trieste, Italy}
\affiliation{The Abdus Salam International Centre for Theoretical Physics (ICTP), Strada Costiera 11, I-34151 Trieste, Italy}
\author{G. Profeta}
\affiliation{Department of Physical and Chemical Sciences, University of L'Aquila, I-67100 L'Aquila, Italy}
\affiliation{SPIN-CNR, University of l'Aquila, I-67100 L'Aquila, Italy}

\date{\today}

\begin{abstract}
We investigated the electronic structure of the Si(111)--7$\times$7 surface below 20 K by scanning tunneling and photoemission spectroscopies and by density functional theory calculations. Previous experimental studies have questioned the ground state of this surface, which is expected to be metallic in a band picture because of the odd number of electrons per unit cell. Our differential conductance spectra instead show the opening of an energy gap at the Fermi level and a significant temperature dependence of the electronic properties, especially for the adatoms at the center of the unfaulted half of the unit cell. Complementary photoemission spectra with improved correction of the surface photovoltage shift corroborate the differential conductance data and demonstrate the absence of surface bands crossing the Fermi level at 17 K. These consistent experimental observations point to an insulating ground state and contradict the prediction of a metallic surface obtained by density functional theory in the generalized gradient approximation. The calculations indicate that this surface has or is near a magnetic instability, but remains metallic in the magnetic phases even including correlation effects at mean-field level. We discuss possible origins of the observed discrepancies between experiments and calculations.
\end{abstract}

\maketitle


\section{\label{sec:intro} Introduction}

Several two-dimensional solid surfaces are considered useful platforms for studying many-body quantum phenomena because the reduced dimensionality makes electronic and structural instabilities more prominent. These instabilities include possible charge and spin density waves \cite{tosatti1974, tosatti1975}, magnetic and Mott metal-insulator transitions\cite{santoro1999}, and superconductivity.
Important examples are represented by some semiconductor surfaces with an odd number of electrons per unit cell, expected to be metallic in a band picture, which have insulating ground states or undergo phase transitions at low temperatures attributed to strong electron correlations or electron-phonon coupling (see $e.g.$ Refs. \onlinecite{carpinelli1996a,carpinelli1997a, weitering1997a,ramachandran1999a, avila1999a, modesti2007a,profeta2007a, cort2013a, hansmann2013a, li2013a,ming2017a, weitering-2019}).  
In this context, the nature of one of the most known and studied surface, the Si(111)--7$\times$7 surface, is still uncertain, with no clear answer about its ground state, metallic or insulating, in spite of the many experimental and theoretical works dealing with it (see $e.g.$ Refs. \onlinecite{Demuth1983a, uhrberg1998a, losio2000a, schillinger2005, barke2006a, modesti2009a, odobescu2015a, ortega1998a, tanikawa2003a} and references therein).

\begin{figure}[ht]
\includegraphics[scale=0.3]{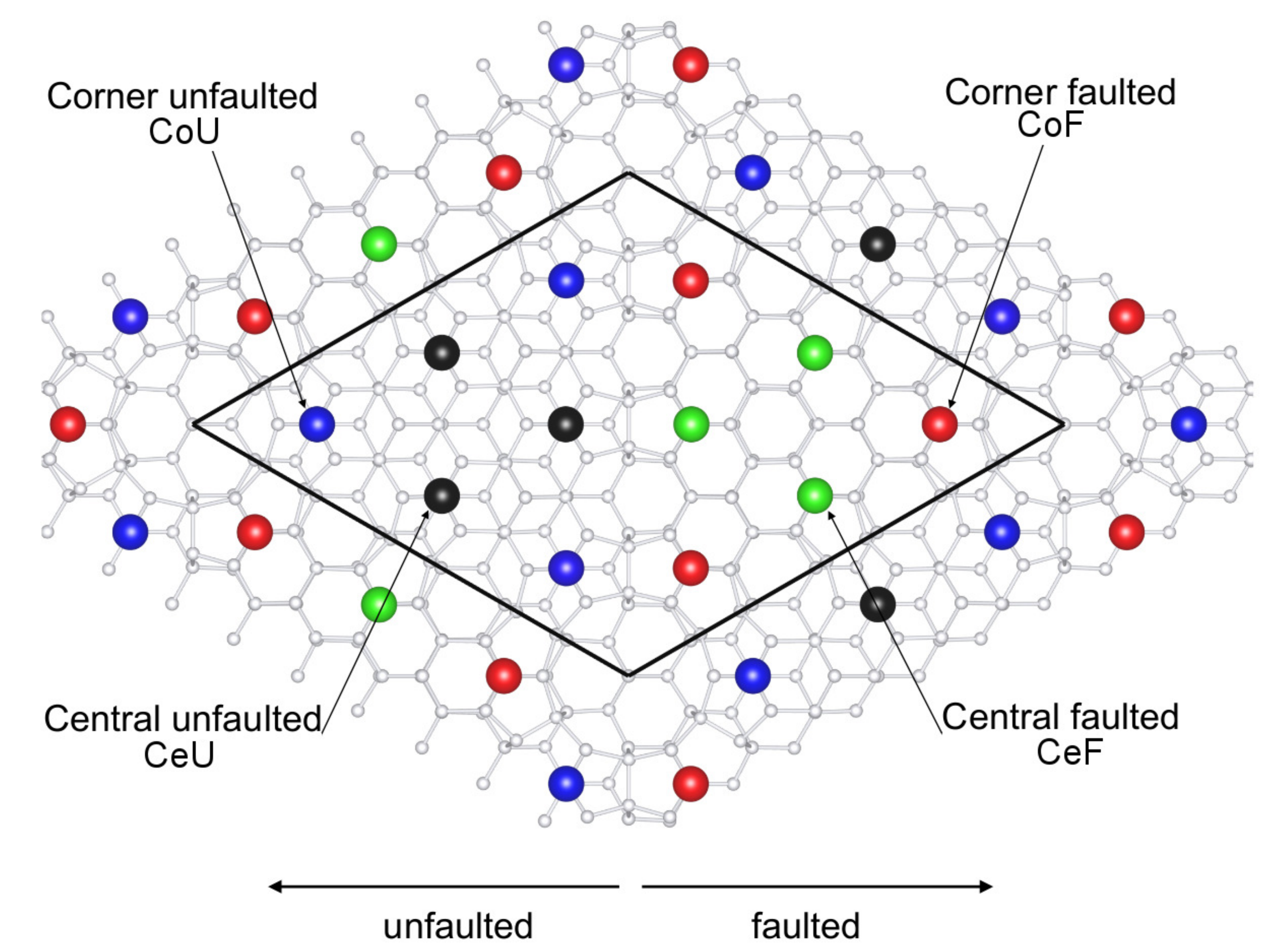}
\caption{\label{fig:struttura}(color online) DFT relaxed structural model of the Si(111)--7$\times$7 DAS surface reconstruction. The unit cell is indicated in black together with  its faulted and unfaulted  halves.
Adatoms and other silicon atoms are indicated by large colored spheres and small gray circles, respectively.}
\end{figure}

The complex Si(111)--7$\times$7 unit cell is generally described by the widely accepted dimer-adatom-stacking-fault model \cite{takayanagi1985a} (DAS model) sketched in Fig. \ref{fig:struttura}, which optimizes the trade-off between strain and broken-bond density to minimize the energy \cite{notaDemuth}. The bulk-truncated Si(111) surface has one broken bond per atom, $i.e.$ 49 broken bonds in the 7$\times$7 unit cell. The 7$\times$7 reconstruction adds 12 Si adatoms of four different kinds and a stacking fault in one of the two triangular halves of the unit cell. The adatoms saturate 36 broken bonds (12 dangling bonds are left on the adatoms) and the stacking fault removes 6 more broken bonds. The 7 remaining broken bonds are located on the 6 surface rest-atoms and in the corner hole at the corners of the unit cell. Density functional theory (DFT) calculations in different approximations show that 7 of the 12 unpaired electrons of the adatoms fill the rest-atom and corner-hole broken bonds \cite{fujita1991a, brommer1993a, ortega1998a, smeu2012a}. 
The 5 remaining electrons partially populate the narrow adatom-derived surface bands placed in the middle of the Si bulk gap (see below). According to the simulations two of these surface bands cross the Fermi level (E$_{F}$)\cite{fujita1991a, ortega1998a}, thus making the surface metallic as expected, in a mean-field approximation, for a non-magnetic system with an odd number of electrons per unit cell. The calculated non-magnetic surface density of states (DOS) of the system has a peak at E$_{F}$ \cite{fujita1991a, smeu2012a} and the charge density near E$_{F}$ is evenly distributed on the four different kinds of adatoms of the unit cell within a factor three \cite{brommer1993a, smeu2012a}.

The metallic picture is consistent with the experimental data measured at room temperature (RT) by photoemission and scanning tunneling spectroscopy (STS) (see $e.g.$ Refs. \onlinecite{uhrberg1998a, martensson1987a, myslivecek2006a}) 
and with some photoemission data at low temperature \cite{Demuth1983a,losio2000a,barke2006a}, but is in contrast with other low-temperature photoemission spectra \cite{uhrberg1998a} and with the hard or soft energy gap at E$_{F}$ observed below $\sim$ 20 K in the STS spectra \cite{modesti2009a, odobescu2012a} and inferred from charge transport experiments \cite{odobescu2015a}. Surface conductivity data show an Efros-Shklowskii ($e^{-1/\sqrt{T}}$) temperature (T) dependence \cite{odobescu2015a}, which is typical of classical two-dimensional disordered systems with a long-range Coulomb interaction among totally localized surface carriers \cite{efros1975a}. This dependence suggests a soft V-shaped Coulomb energy-gap with vanishing single-particle DOS at E$_{F}$ at low temperature, caused by the long-range interactions \cite{odobescu2015a, shklovskii, efros1975a}. The hard gap, tens of meV wide, observed in the tunneling spectra \cite{modesti2009a, odobescu2012a} results from the soft Coulomb gap modified by a local Coulomb-blockade effect according to Ref.  \onlinecite{odobescu2015a}. Additional hints or indications of a ground state different from that of a normal metal have been known for many years and are provided by high-resolution energy-loss spectroscopy \cite{Demuth1983a, persson1984a} and nuclear magnetic resonance experiments \cite{schillinger2005, fick2006} at low temperature. These data have been interpreted in terms of an extremely narrow  band at E$_{F}$ within a band gap of the surface states. Other clues are provided by transport experiments \cite{tanikawa2003a, wells2006a, dangelo2009a, martins2014a, just2015a}, which measure a surface conductance too low for a normal metal, $i.e.$ well below the two-dimensional Joffe-Regel limit \cite{dangelo2009a}.

Electron correlation is a possible explanation of the above-mentioned experimental results mainly because the screened on-site Coulomb interaction on the Si adatoms, estimated to be about 1 eV \cite{ortega1998a, hansmann2013b}, is larger than the energy span of the calculated adatom dangling-bond bands near E$_{F}$. A first calculation by Ortega {\em et al.} \cite{ortega1998a} with a model that includes strong correlation effects points out their importance, although their simulation predicts a metallic state. Correlation effects have also been proposed to explain the width of the adatom bands observed by photoemission \cite{losio2000a}.  In addition, strong electron-phonon coupling, found according to the analysis of the photoemission spectra presented by Barke {\em et al.} \cite{barke2006a} and the computed phonon dispersions of group IV adatoms on Si(111)\cite{perez2001a}, can alter the bands calculated by DFT in the local density approximation (LDA). 
Another element that could affect the surface bands is the possible existence of magnetic phases\cite{sheka2003, mag-silicon}.

In principle, photoemission spectroscopy and STS provide direct access to  the low-temperature electronic structure of surfaces. However, specific difficulties arise in the interpretation of the data in the case of Si(111)--7$\times$7 that explain the conflicting results of the literature. The surface photovoltage (SPV) effect\cite{alonso1990a, marsi1998a} introduces in the low-temperature photoemission spectra an indeterminacy of the E$_{F}$ position of at least several tens of meV. This energy scale is critical for the Si(111)--7$\times$7 surface and makes different datasets compatible with a metallic\cite{Demuth1983a, losio2000a, barke2006a} or insulating\cite{uhrberg1998a} ground state.  Low-temperature STS data\cite{modesti2009a, odobescu2012a, odobescu2015a, myslivecek2006a, wang2018a} show the opening of a hard or soft gap below 20 K. However, the published STS spectra are integrated over the unit cell, or taken only on some kinds of the adatoms, or not suitably corrected for the systematic error caused by the non-equilibrium transport of carriers, which affects the energy scale at low temperatures\cite{modesti2009a, myslivecek2006a}. Hence, these analyses do not allow us to identify the adatom-dependent effects of a gap opening at E$_{F}$.

In order to clarify the low temperature state of the Si(111)--7$\times$7 surface, we performed temperature-dependent photoemission and STS experiments. Thanks to our improved correction of the SPV the photoemission spectra clearly show that no band crosses the Fermi level at 17 K, thus excluding a metallic ground state. We also measured accurately corrected low-temperature tunneling spectra resolved on the four kinds of adatoms in the unit cell. Both photoemission and tunneling spectra consistently indicate a strong reduction of the spectral intensity near E$_{F}$ from RT to 17 K.  Adatom-selective STS spectra show that this spectral intensity decrease also involves all the filled dangling bond states of one of the four kinds of adatoms. As a result, while at room temperature metallic surface states near E$_{F}$ are spread over all adatoms in the cell, a more inhomogeneous evolution takes place at low temperatures, where the weight of insulating states just below E$_{F}$ concentrates in three corner faulted adatoms closest to the vacancy site, surprisingly dropping in some of the others. We attempted to explain the observed insulating state by performing state-of-the-art DFT calculations in the generalized gradient approximation (GGA) of the structural and electronic properties of the Si(111)--7$\times$7 surface. We allowed both the spin degree of freedom and the breaking of the C$_{3v}$ symmetry of the DAS model, and included strong electron correlation effects at the level of the DFT+U approximation.  The calculations indicate that, while
magnetic and non-magnetic phases,
as well as undistorted or weakly Jahn-Teller distorted phases,
are all nearly degenerate on this surface,
no clear instability mechanism of the metallic state appears in this approximation. 
These results support the view that the actual mechanism that drives the insulating state and the massive spatial redistribution of spectral density  at low temperature must be of non-mean field nature.

\section{Experimental and Computational Methods}

The samples used for the STS measurements were heavily $n$- (As-) doped Si(111) wafers with a resistivity  of about 0.005 $\Omega$ cm. The photoemission spectra were acquired at the VUV-Photoemission beamline of the Elettra synchrotron radiation source also from $p$- (B-) doped (1-20 $\Omega$ cm) and undoped (\textgreater 2000 $\Omega$ cm) samples. The surfaces were cleaned by flash heating up to 1600 K in ultra-high-vacuum (UHV) by electron bombardment and annealed at about 1100 K to obtain a well-ordered 7$\times$7 reconstruction. The total energy resolution (photon + electron) of the photoemission spectra was 15 meV full width at half maximum (FWHM). The polarization of the synchrotron light was horizontal and laid in the scattering plane of the experiment, defined by the incoming photon beam and the axis of the cylindrical lens system of the electron analyzer.  The sample temperature was estimated to be 17$\pm$3 K by replacing the sample with a Si-diode thermometer. The low-temperature spectra were corrected for the SPV shift \cite{alonso1990a, marsi1998a}  within $\pm$15 meV  as described in the Appendix A. The SPV effect also induces an energy broadening of the spectra due to the spatial inhomogeneity of the photon beam and the sample and the pulsed nature of the photon flux \cite{marsi1998a}.

STS spectra and differential conductance (dI/dV) maps were measured far from defects on 7$\times$7-reconstructed terraces with low surface defect densities (\textless 0.5 10$^{16}$ m$^{-2}$, about 1 defect per 30 7$\times$7 unit cells) by using a home-made STM with a gold tip. The STS spectra at 7 K have been corrected for the systematic errors caused by the transport of the charge carriers from the surface to the sample holder by using the method described in Ref. \onlinecite{modesti2009a} with a tunneling current less than 1 pA between -0.5 and 0.5 V.  This method allows the separation of the total bias voltage  V$_{bias}$ applied in STS in the potential drop across the tunnel junction between the tip and the surface region just below it, and the potential drop between this region and the sample holder.  The surface and bulk carrier transport affects the latter potential drop, not the former. In this work the values of the voltage V used to compute and to plot the spectra at 7 K are, correctly, relative to the potential drop across the tunnel junction and not to the total drop between the tip and the sample holder \cite{modesti2009a}. At 70 K and 295 K the STS spectra were measured with currents less than 1 pA and did not need the corrections mentioned above. The comparison between STS spectra acquired at different temperatures could be affected by accidental modifications of the tip and of its DOS during the temperature change. For this reason every STS spectrum reported in this article is the average of at least ten spectra obtained with different Au tips. We assume that the average electronic structure of the Au tips does not depend on temperature between 7 and 300 K. Under this assumption we consider the observed spectral changes as a function of T as due to variation of the electronic structure of the sample.
 At each temperature sets of STS spectra were acquired on the four different types of adatoms cyclically. Modifications of the tip structure were intentionally obtained by touching the Si surface between the acquisition cycles. As shown in appendix B on data reproducibility and data analysis, the 
 characteristic spectral features of the different kinds of adatoms and the temperature dependence of the spectra do not depend on the tip properties. The maximum broadening in the dI/dV spectra due to numerical differentiation of I(V) is 0.01 V near the gap region.

Maps of the average dI/dV between -0.07 V and -0.20 V were obtained by acquiring I-V spectra on a 32$\times$32 spatial grid and computing the increment ratio of the current with respect to the bias voltage. The STS spectra and the conductance maps were acquired stabilizing the tip position at a bias voltage of 1.2 V or 1.4 V with a tunneling current of 3 pA or less. In this bias range the maximum spread of the apparent height of different adatoms in the STM topographic images is 0.15 \AA. Taking into account the vertical position of the adatoms predicted by ab-initio calculations \cite{smeu2012a,ke2000a} the vertical tip-adatoms distance did change by less than 0.25 \AA~ from one type of adatom to another. 
 
First-principles DFT calculations were performed within the pseudopotential approximation\cite{Bloch_PAW_PRB1994} with the Vienna ab-initio Simulation package (VASP)\cite{Kresse_VASP_PRB1996, Kresse_VASP_CompMatSci1996}, in the GGA approximation\cite{GGA-PBE}. 
The Si(111)--7$\times$7 surface was constructed according to the DAS model with C$_{3v}$ symmetry, with three Si bilayers to simulate the bulk-like substrate. The kinetic energy cutoff was set to 400 eV and the $k$-point sampling of the Brillouin zone to a 5$\times$5 Monkhorst-Pack uniform mesh\cite{M_and_P_kmesh_PRB1976}. The bottom Si surface was saturated with hydrogens. Structural relaxation of the surface layer and of the topmost two Si bilayers was performed up to residual forces of 0.01 eV/\AA.
DOS and projected density of states (PDOS) were evaluated with $k$-point mesh up to 15$\times$15 and electronic smearing of 0.01 eV.

\section{Results and analysis} 
\subsection{DFT electronic structure - non-magnetic phase}
\label{theory}
\begin{figure}[h]
\includegraphics[scale=0.55, trim=1.3cm 0 0 -0.cm]{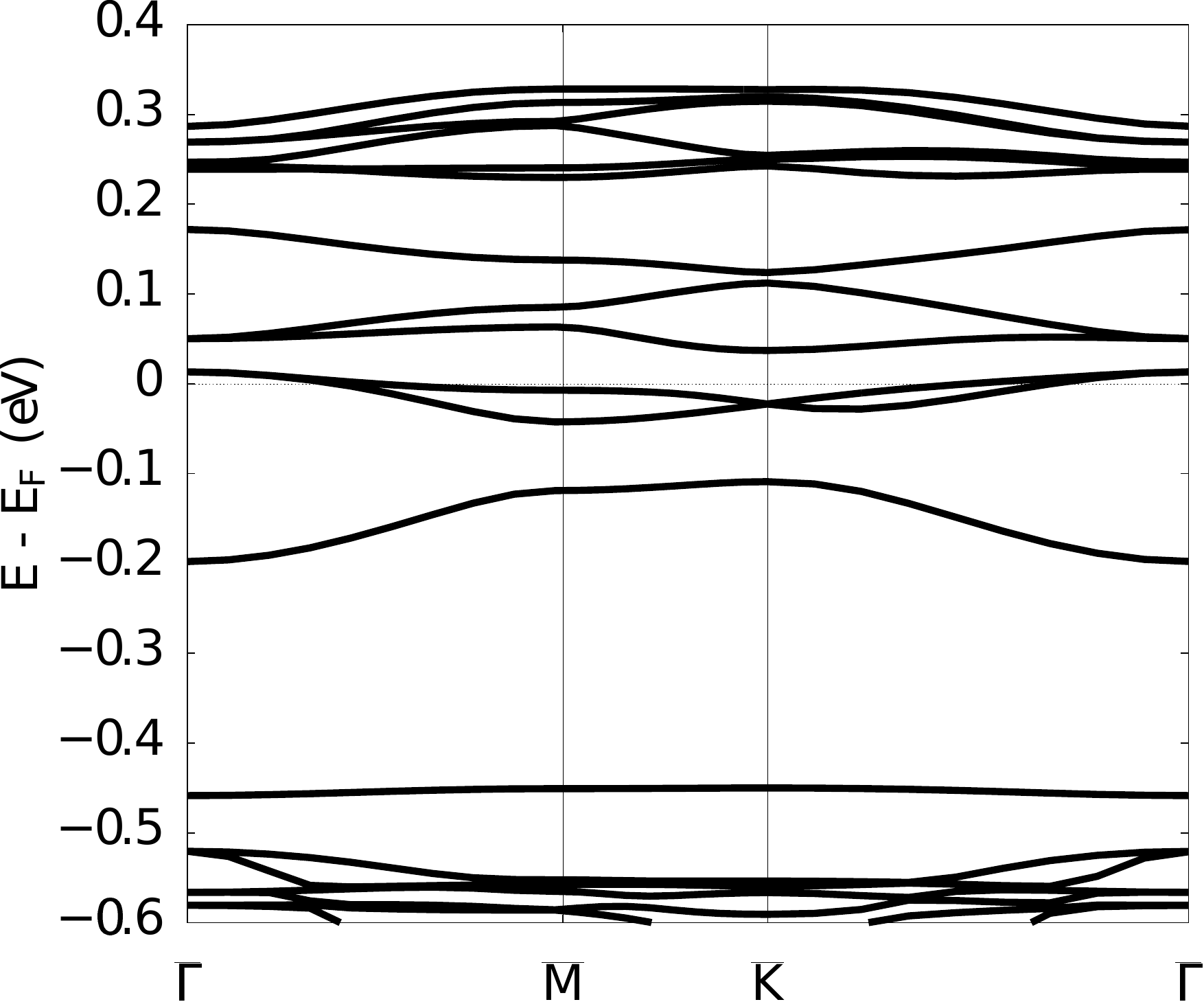}
\includegraphics[scale=0.65]{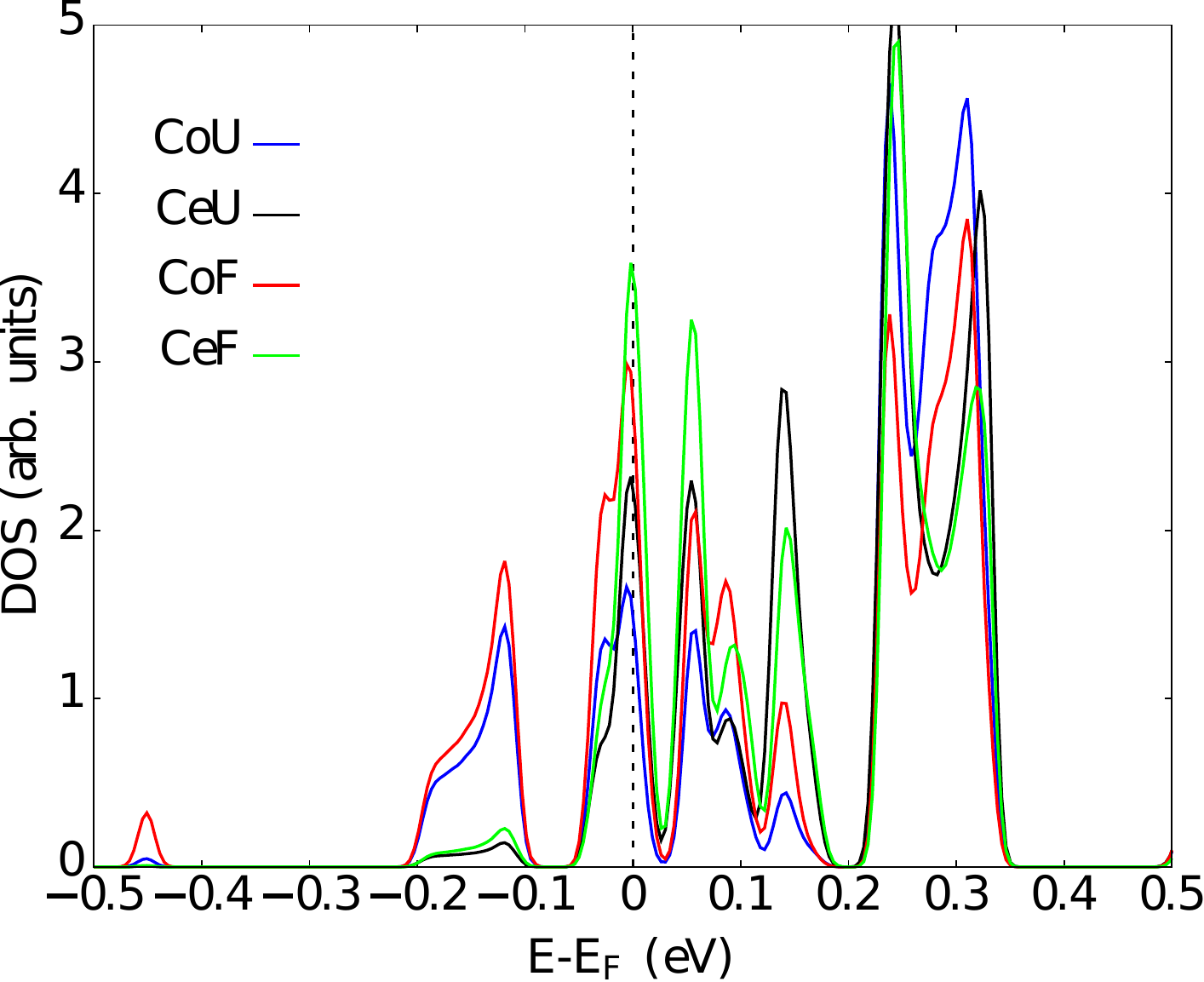}
 \caption{\label{fig:bands_normal} (Upper panel) Band structure of the non-magnetic Si(111)--7$\times$7 surface and (lower panel) projected density of states on the adatom sites: corner (blue) and central (black) adatoms in the unfaulted region and  corner (red) and central (green) adatoms in the faulted region.}
\end{figure}

Fig.\ref{fig:bands_normal} shows the dispersion of the electronic bands close to E$_{F}$ calculated for a non-magnetic Si(111)--7$\times$7 surface. We find 12 surface bands deriving from the 12 adatoms dangling bonds between -0.2 and 0.4 eV in the gap of the Si bulk states. The flat band at -0.45 eV arises mainly from the surface atom at the center of the hole. Projected bulk states are present below -0.5 eV.

The surface is predicted to be metallic, as expected for a single-particle band picture, with two narrow bands crossing the Fermi level in the region between the center and the edges of the 7$\times$7  surface Brillouin zone (SBZ), in agreement with the results of Ortega $et$ $al.$ \cite{ortega1998a} 
In order to compare the DFT results with the experimental photoemission and dI/dV spectra the adatom PDOS and the local density of states (LDOS) 3 \AA\ and 5 \AA\ above the surface were computed for each kind of adatom.  
The calculated PDOS, presented in Fig. \ref{fig:bands_normal}, confirms that the states close to E$_{F}$ are delocalized over all the adatoms in the surface unit cell. The PDOS at and below E$_{F}$ of the adatoms in the faulted part is higher than that of the adatoms in the unfaulted part. In each half, the central adatoms have a higher PDOS than the corner adatoms, in agreement with other computational results\cite{smeu2012a}. As shown in Fig. \ref{fig:bands_normal} the PDOS has a relative maximum very close to E$_{F}$ on all the adatoms, related to an inflection point of the band dispersion at the $\overbar{\textrm{M}}$ point (see Fig. \ref{fig:bands_normal}), and the values of the PDOS on inequivalent adatoms differ by less than a factor 2.5 in the first 0.1 eV below E$_{F}$. Low energy states centered at -0.15 eV derive mainly from the corner adatoms. The small peak at about -0.45 eV is due to the interaction between the CoF adatoms and the atom at the center of the hole.
The shapes and relative weights of the LDOS near E$_{F}$ calculated 3 \AA~ and 5 \AA~ above each adatom (not shown)  do not depend on the distance from the surface and are similar to those of the PDOS within 20\%, indicating that, in the Tersoff-Hamann approximation \cite{tersoff1985a}, the experimental differential conductance on different adatoms (see below) can be interpreted, at least qualitatively, by means of the relative PDOS near E$_{F}$.

\subsection{Scanning tunneling spectroscopy}
\label{sec:sts}
\begin{figure}[ht]
\includegraphics[scale=0.15]{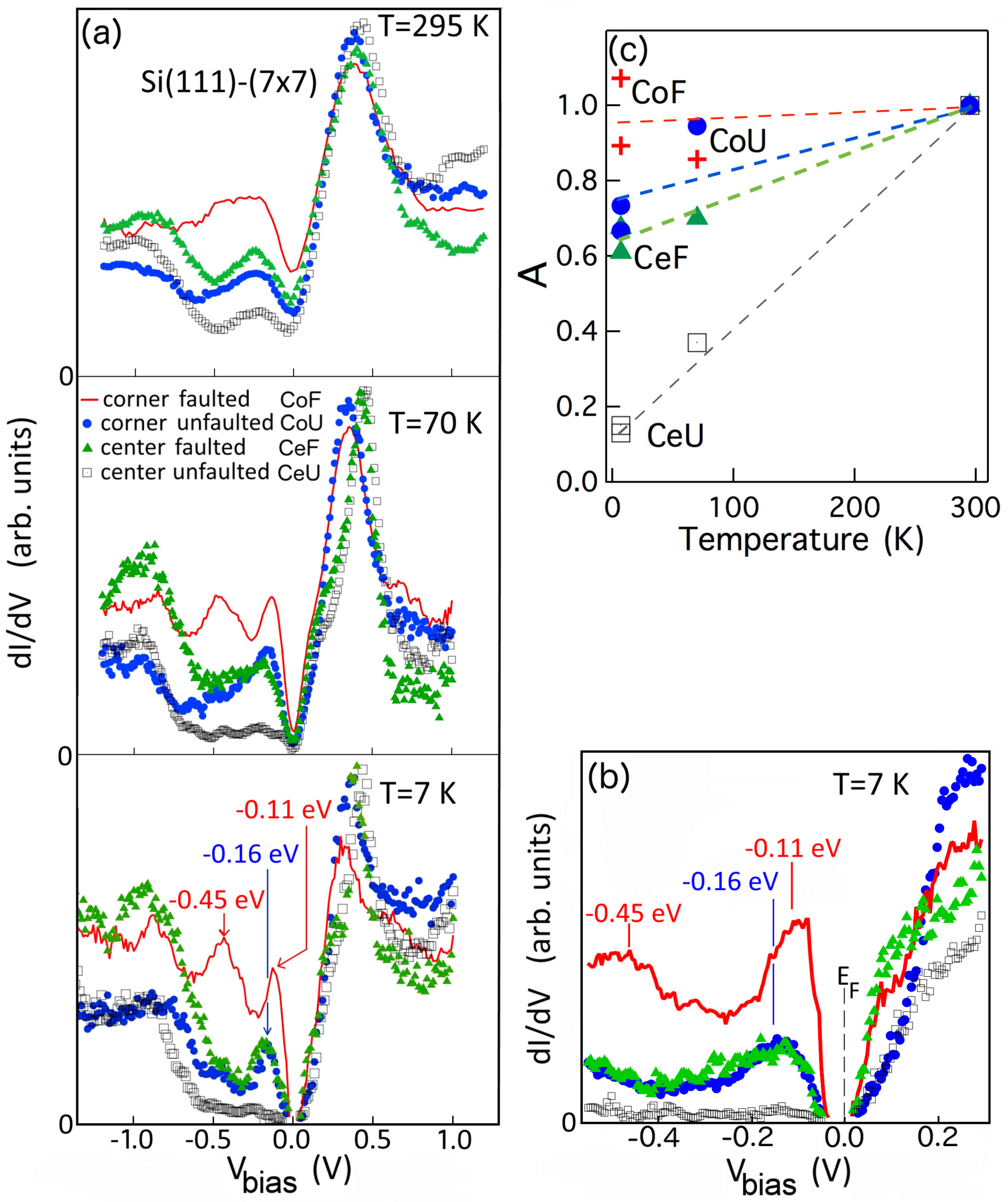}
\caption{\label{fig:sts} (color online). (a) Tunneling spectra measured on top of the four kinds of adatoms in the metallic phase at 295 K and 70 K and in the insulating phase at 7 K keeping the tip-adatom distance constant (stabilization setpoint V$_{bias}$=1.2 V and I$_{tunneling}$=3 pA). (b) Spectra in the energy region close to the Fermi level at 7 K. The spectra of this panel are from a set of data different from that of panel (a) at 7 K obtained with different tips on a different sample. (c) Temperature dependence of the averages $A$ of dI/dV from -0.20 V to -0.07 V of different adatoms. $A$ is normalized to 1 at 295 K for all the adatoms. Dashed lines are guides for the eye.}
\end{figure}

Fig. \ref{fig:sts} shows the temperature dependence of dI/dV of the four types of adatoms of the 7$\times$7 unit cell. All spectra at 7 K (lower panel of Fig. \ref{fig:sts}(a) and Fig. \ref{fig:sts}(b)) have energy gaps at E$_{F}$ of similar width (about 0.05 eV),  indicating an insulating phase. This is in agreement with previous results obtained by averaging the spectra over all the adatoms\cite{modesti2009a} or only some \cite{odobescu2012a, odobescu2015a} of the adatoms in the unit cell, but contrasts with the calculated PDOS, which shows a maximum at E$_{F}$. The two sets of data taken at 7 K on different samples with different tips show the degree of reproducibility of our tunneling spectra. The data noise in the region near the gap sets the upper limit of dI/dV at E$_{F}$ to ~6\% of the value measured at 295 K, $i.e.$ the ratio R$_{STS}$ of dI/dV at 7 K to that at 295 K at E$_{F}$ is 0.03$\pm$0.03. More details on the gap region of the tunneling spectra are reported in the second appendix. The spectra at 70 K show finite dI/dV values at E$_{F}$ for all the adatoms and increased spectral intensity for the CeU adatoms near the gap with respect to the 7 K case.

\onecolumngrid

The data of Fig. \ref{fig:sts} indicate a new aspect of the low-temperature electronic properties, namely the low dI/dV values on CeU adatoms in the first few hundreds mV below E$_{F}$ at 7 K, which are one order of magnitude lower than those of the other adatoms. 
The drop of the spectral intensity at low temperature is quantified in Fig. \ref{fig:sts}(c), which reports the mean value of the dI/dV  between -0.07 and -0.20 V ($A$) normalized to its value at 295 K as a function of the temperature. $A$ decreases by more than a factor 6 from 295 to 7 K for the CeU adatoms, and by less than a factor 2 for the other adatoms. This adatom-specific behavior rules out spurious tip effects and highlights temperature-dependent changes of the electronic structure that affect the PDOS of the CeU adatoms near -0.2 eV or/and their tunneling probability\cite{feenstra1987a, tersoff1985a, notasts}.   
Notably, DFT calculations of Section \ref{theory}
of the metallic, nonmagnetic and undistorted DAS surface do not explain the experimental observation, at least in the Tersoff-Hamann approximation.\cite{tersoff1985a} Adatom-resolved PDOS and LDOS differ by less than a factor 2.5 near E$_{F}$, while dI/dV measured on the CeU adatoms is one order of magnitude smaller than on the other adatoms at 7 K.

In addition to the experimental insulating gap on all the adatoms, Fig. \ref{fig:sts} shows well-defined peaks at 7 K between -0.11 V and -0.45 V, which correspond to features observed in the photoemission data of Refs. \onlinecite{uhrberg1998a, barke2006a} and, therefore, can be associated to maxima of the surface DOS. 
The peak at -0.45 V on the CoF adatoms is also present in the calculated PDOS on the same adatom (see Fig. \ref{fig:bands_normal}). It originates from the interaction with the dangling bond of the atom at the center of the hole. 
The peaks of the topmost filled states of the CeF and CoU adatoms, centered at -0.16 V, are slightly shifted with respect to that of the CoF adatom at 0.11 V (see Fig. \ref{fig:sts}(b)).  The other peaks, already reported in previous studies at higher temperatures \cite{myslivecek2006a, wolkow1988a}, are the maximum at -0.9 V, which is degenerate with the peak on the rest-atoms \cite{brommer1993a, smeu2012a, wolkow1988a}, and the peak of the empty adatom states in the range of 0.2-0.4 V.

\begin{figure}[ht]
\includegraphics[scale=0.17 ]{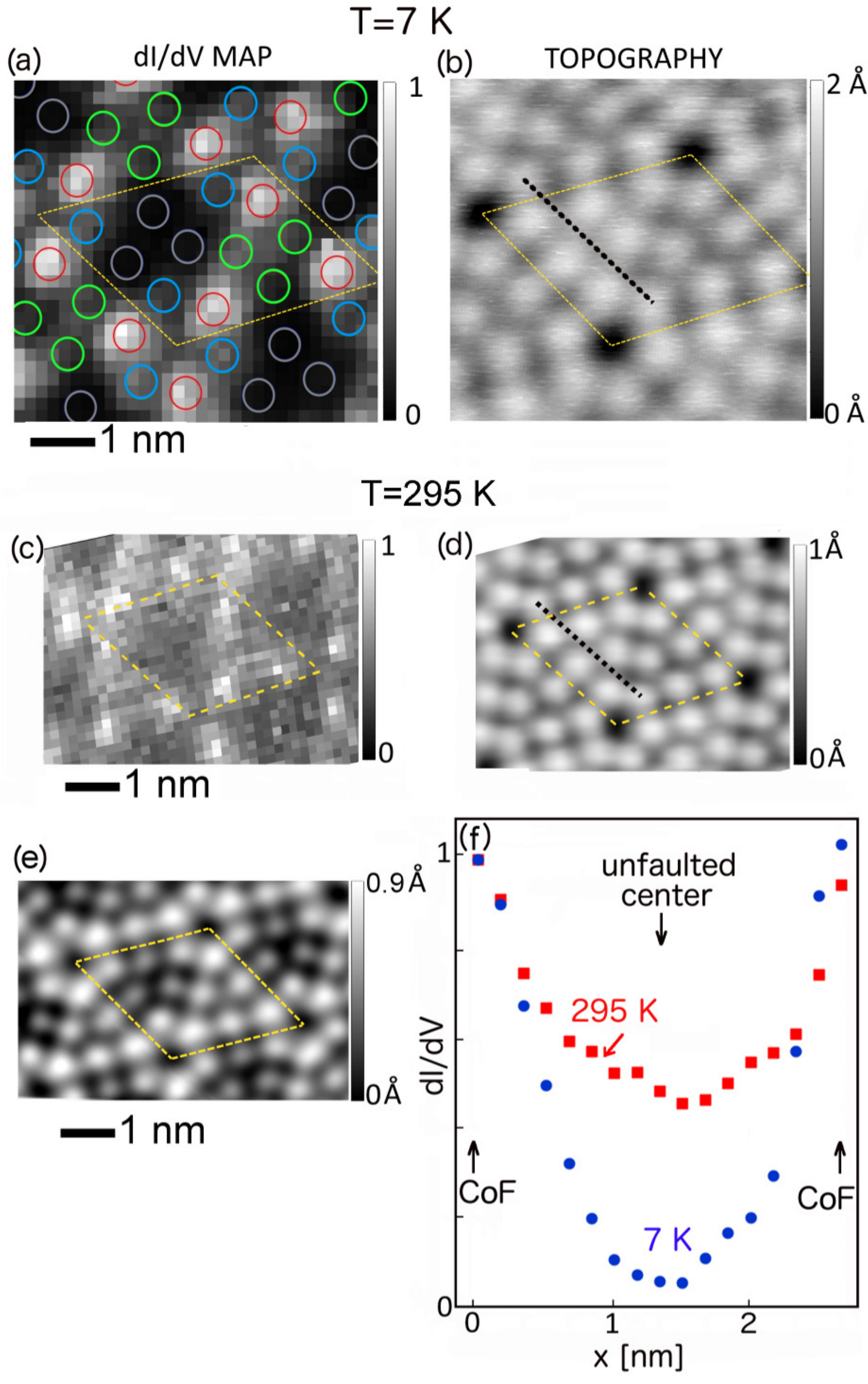}
\caption{\label{fig:sts-map} (color online).  (a) Differential conductance map at 7 K in the region shown in the topographic map in panel (b). The stabilization set point was V$_{bias}$=1.4 V and I$_{tunneling}$=5 pA both in spectroscopy and in topography. The reported dI/dV values are the average values between -0.07 and -0.20 V and are normalized to the maximum average value measured on the CoF adatoms. The positions of the four kinds of adatoms are marked by circles, the dashed parallelogram indicates the unit cell. The images are not corrected for the geometrical distortion caused by the drift of the sample. Panels (c) and (d) are the same as (a) and (b) but at 295 K. The stabilization set point was V$_{bias}$=1.4 V and I$_{tunneling}$=50 pA both in spectroscopy and in topography. The larger effect of the drift or the creep of the piezo and the different quality of the 7 K image with respect to that at 295 K are caused by the much longer acquisition time for the dI/dV map at 7 K due to the very low tunneling current used at low temperature to minimize the charge transport effects. (e) Topographic image acquired with a negative V$_{bias}$=-0.4 V and I$_{tunneling}$=2 pA at 7 K. (f) Intensity profiles of dI/dV at 295 K and 7 K measured along the black dashed lines in panels (b) and (d), which connect two CoF adatoms passing through the center of the unfaulted region. The CoF adatoms are at 0.0 nm and at 2.68 nm, the center of the unfaulted zone is at 1.34 nm. The larger effect of the drift or the creep of the piezo and the different quality of the 7 K image with respect that at 295 K are caused by the much longer acquisition time for the dI/dV map at 7 K due to the very low tunneling current used at low temperature to minimize the charge transport effects.”}
\end{figure}

Fig. \ref{fig:sts-map}(a) shows the dI/dV map at 7 K averaged between -0.07 V and -0.20 V. This map confirms that the distribution of dI/dV  related to the filled states close to the gap is strongly peaked on the CoF adatoms, while the regions of the CeU adatoms are empty, with a ratio equal to $\sim$25  between maxima and minima of the averaged dI/dV.  This ratio decrease to 2 in the dI/dV map at 295 K (Fig. \ref{fig:sts-map}(c) and (e)), in line with the room temperature STS spectra of Fig.  \ref{fig:sts}. These data support the distinct temperature dependence of the electronic properties of the CeU adatoms. 

A map qualitatively similar to that of Fig. 4(a) is obtained at negative stabilization bias voltages. The main difference is a decrease of the contrast by approximately a factor 3 caused by the reduction of the tip-adatom distance in the unfaulted half of the unit cell necessary to compensate for the smaller PDOS of the occupied states. As shown in Fig. 3 the dI/dV signals differ by at most a factor 2 just above the gap, and are larger on the faulted half of the unit cell. Therefore, we expect that the dI/dV map of the empty states close to the gap  displays a substantially smaller contrast,  and again a minimum in the unfaulted part. This does not support a charge density wave picture of the ground state.

\subsection{Angle-integrated photoemission} 

In order to verify the insulating state observed by the STS experiments, we performed a complementary investigation of the low-temperature electronic structure of Si(111)--7$\times$7 by photoemission spectroscopy. 
\begin{figure}[ht]
\includegraphics[scale=0.15]{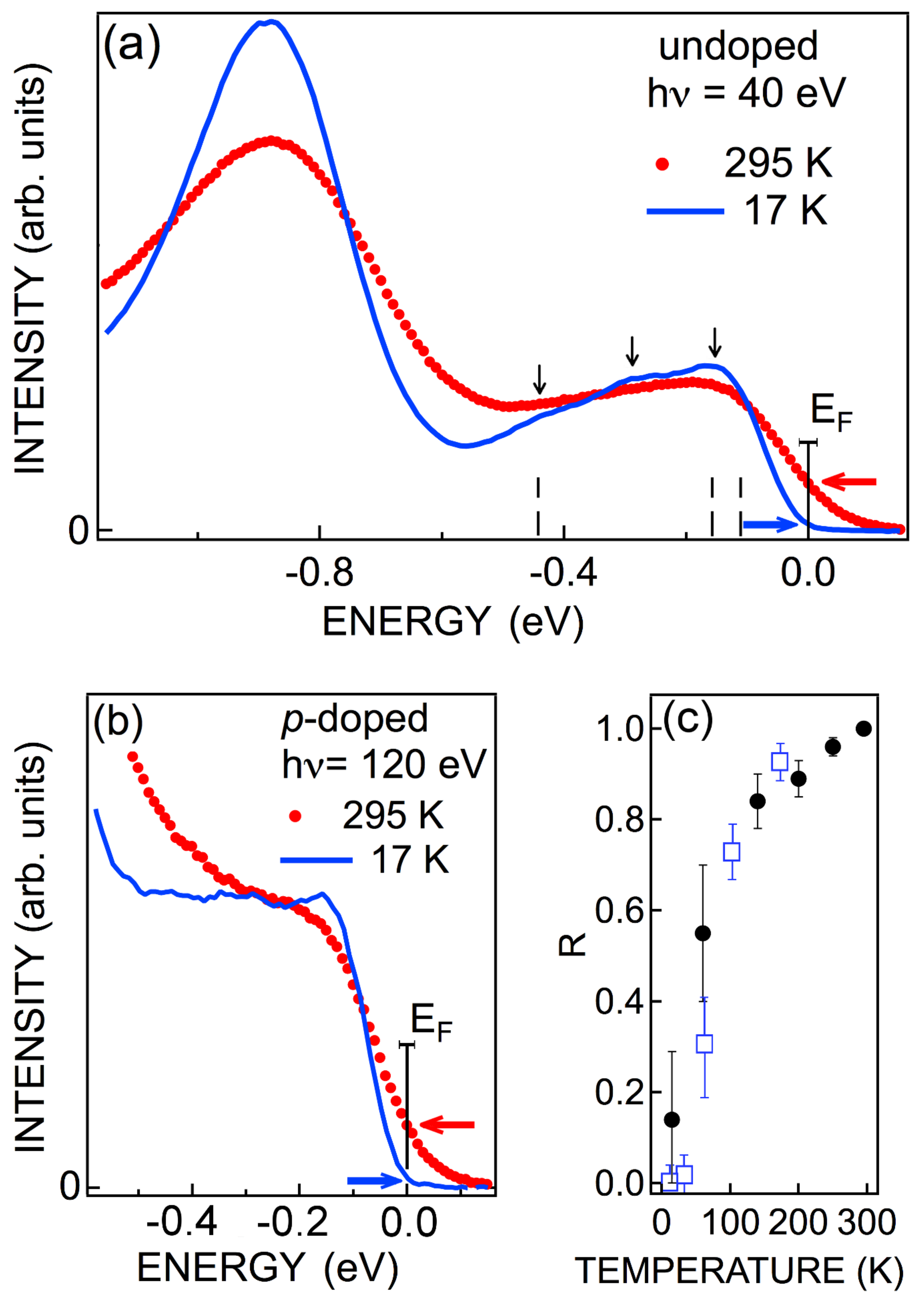}
\caption{ (color online). Angle-integrated photoemission spectra measured at 295 K and 17 K on an undoped sample using a photon energy of 40 eV (a) and on a $p$-doped sample using a photon energy of 120 eV (b). The vertical solid line marks the Fermi level. The horizontal error bar on this line indicates the maximum error of E$_F$ at 17 K caused by the SPV shift (see appendix A). The horizontal arrows indicate the spectral intensities at E$_F$ at the two temperatures. The vertical dashed lines indicate the energy positions of the STS adatom peaks, while the vertical arrows point to the photoemission features in the energy window of the adatom dangling bonds. The strong peak at -0.9 eV is caused by the restatom states. (c) Ratio R of the photoemission intensities at E$_F$ at different temperatures to that at 295 K as a function of the temperature (dots) compared with a similar ratio from dI/dV at E$_F$ (squares). The maximum errors of the photoemission data are due to the uncertainty of the position of E$_F$.}
\label{fig:PE}
\end{figure}
The angle-integrated photoemission spectra of Figs. \ref{fig:PE}(a) and \ref{fig:PE}(b), measured at 17 K and corrected for the SPV shift, show strongly reduced intensities at E$_{F}$ with respect to the 295 K data (horizontal arrows), providing a confirmation of the STS results.  
This drop can be quantified by the ratio R of the photoemission intensity at E$_F$ at a given temperature to that at 295 K, reported in Fig. \ref{fig:PE}(c) and compared with a similar ratio from the STS data. The differential conductance drops below the noise level at about 20 K, in agreement with Ref. \onlinecite{odobescu2015a}. At 17 K R from photoemission is 0.15$\pm$0.15, where the error is mainly caused by the propagation of the maximum error of the position of E$_{F}$ in the spectra, which reaches $\pm$15 meV at 17 K. The value 0.3 is the upper limit of the  ratio between the DOS at E$_{F}$ at the two temperatures because the SPV-induced extrinsic broadening of the low-temperature spectra inflates the high-energy tail of the spectra at and above E$_{F}$. The presence of dominant SPV broadening at 17 K is testified by the extent of the high-energy tail (more than 30 meV above E$_{F}$), which is much larger than the experimental resolution (15 meV FWHM) and the thermal broadening of the Fermi distribution ($\sim$6 meV).
To obtain a more accurate estimate of the DOS at E$_F$ we took into account the SPV broadening by fitting the  spectra at 17 K between -0.07 eV and 0.10 eV by the convolution of a Gaussian function,  which describes the broadening, and the function $P(E)=DOS(E)f(E)$, where $E$ is the energy and $f(E)$ is the Fermi distribution at T=17 K. We approximated $DOS(E)$ near E$_F$ by a linear function. One of the free parameters of the fit is the energy of the real Fermi level (that we call $E'_F$) because of its 15 meV maximum error. The other three parameter are $DOS(E'_F)$, the value of the linear function at $E'_F$ normalized to its value at 295 K, the slope of the linear function, and the FWHM of the photovoltage broadening. 

\begin{figure}[ht]
\includegraphics[scale=0.19]{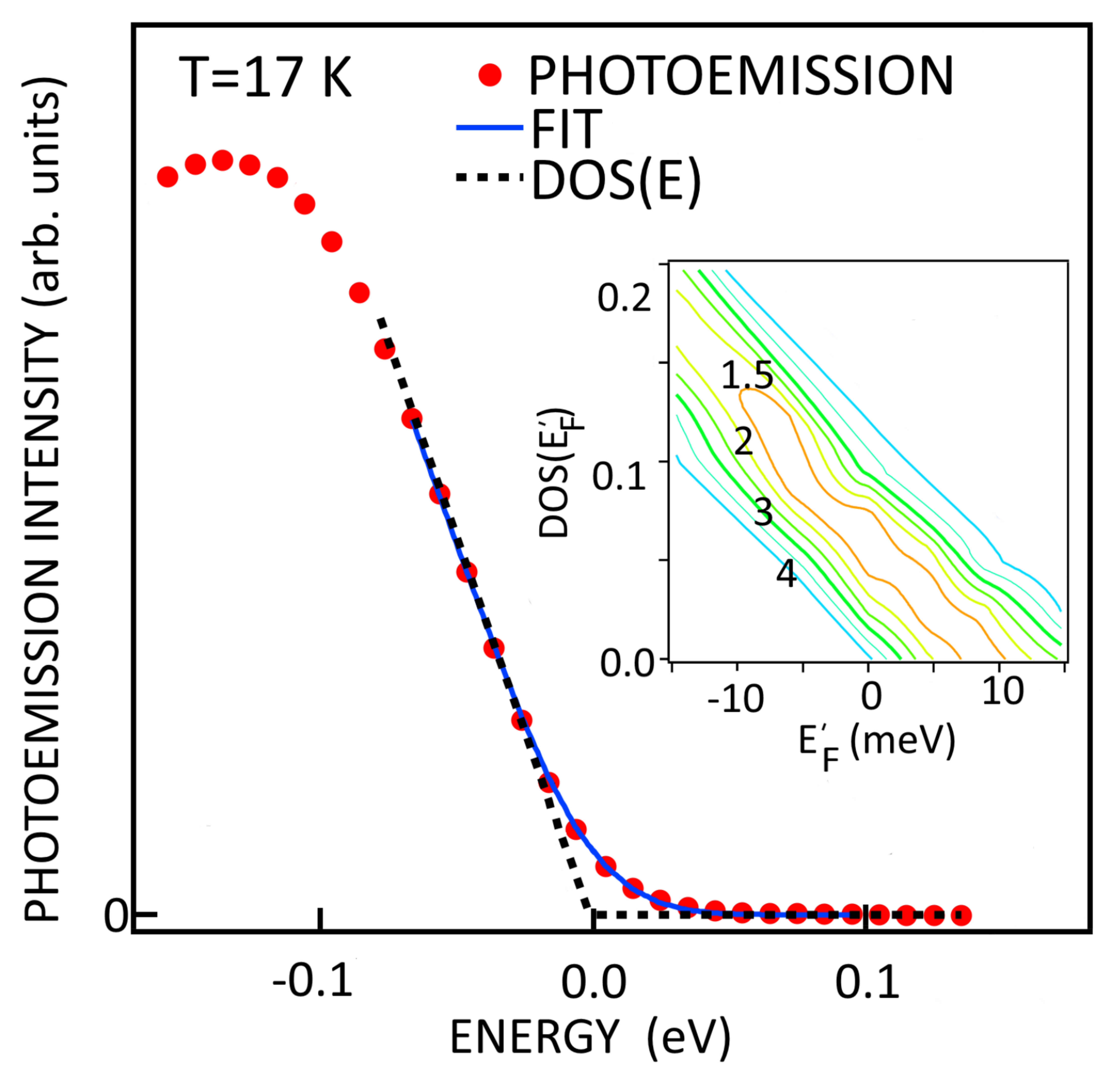}
\caption{\label{fig:PE-fit} (color online). Example of the fit of the Fermi energy spectral region of the angle integrated photoemission spectrum that takes into account the SPV broadening (see text). In the particular case shown DOS(E) was set to vanish at the E$_{F}$. The spectrum (dots) was measured with a photon energy of 40 eV at 17 K on a $n$-doped sample. The zero of energy in this figure is $E'_F$. 
Inset: two-dimensional contour plot of the normalized $\chi^2$  as a function of $DOS(E'_F)$ and $E'_F$ (see text). }
\end{figure}

An example of the fit is shown in  Fig. \ref{fig:PE-fit} for the case of 
 $DOS(E'_F)=0$ and $E'_F=8$ meV \cite{notaEF}, together with the $\chi^2$ (normalized to its minimum value), which measures the quality of the fit as a function of $DOS(E'_F)$ and $E'_F$. The best estimate for the $DOS(E'_F)$ at 17 K is $DOS(E'_F)\in[0,0.2]$. This range is determined by the condition $\chi^2< 2$ and is a consequence of the weak dependence of the quality of the fit on $E'_F$. The FWHM of the SPV broadening is between 0.045 eV and 0.055 eV, confirming that it is considerably larger than the experimental energy resolution and the thermal broadening of the Fermi distribution at 17 K.
The lower limits of both $DOS(E'_F)$ and the corresponding ratio $R_{STS}$ obtained from the STS data (see Sect. \ref{sec:sts}) are zero, consistent with the presence of an energy gap. The upper limit of $DOS(E'_F)$ is higher than that of $R_{STS}$ (0.06) because of the propagation of the SPV-related error and the proximity of the temperature (20-30 K) at which the gap closes \cite{modesti2009a, odobescu2015a}.

The observation of the dip at E$_F$ in the low-temperature photoemission data indicates that the deep minimum seen by STS is caused by a real drop of DOS and not mainly by a local Coulomb blockade, as suggested by Ref. \onlinecite{odobescu2015a}, or temperature-dependent tunneling probability effects, which do not affect the photoemission spectra. 
In addition, the energies of the adatom peaks observed in our STS spectra (indicated by dashed lines in Fig. \ref{fig:PE}(a)) agree with those of the peak at -0.14 eV and the shoulder at -0.45 eV of the photoemission spectrum, thus supporting the assigment of the STS features to maxima of the surface DOS.
The spectra of undoped (Fig. \ref{fig:PE}(a)), $n$-doped (Fig. \ref{fig:PE}(b) and Fig. \ref{fig:PE-fit}) and $p$-doped (Fig. \ref{fig:ARPES}(a), see below) samples present the same low-energy features within the experimental error, thus excluding observable effects of bulk doping, such as the modification of the population of the surface states and the contribution from bulk carriers to the photoemission spectra. 

The two techniques, STS and angle-integrated photoemission, offer in conclusion a similar and coherent picture of a genuine gap-opening metal-insulator transition upon cooling. Alas, they do not explain why and how that transition, not predicted by DFT calculations, actually takes place.  

\subsection{Angle-resolved photoemission} 
\label{sec:ARPES}

\begin{figure}[h]
\includegraphics[scale=0.6]{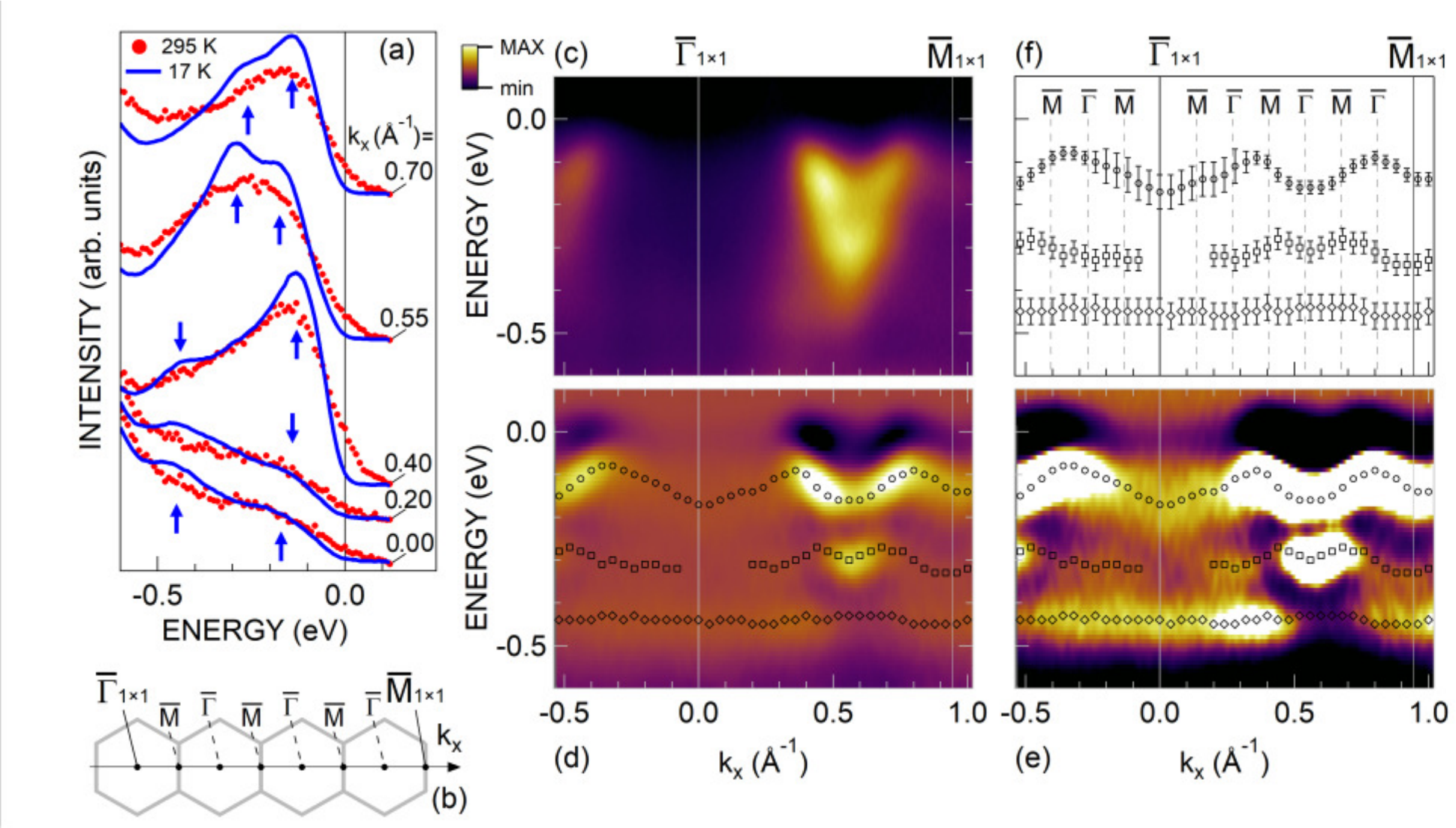}
\caption{\label{fig:ARPES}(color online). (a) Spectra of a $p$-doped sample at $k$=0.00, 0.20 , 0.40, 0.55, and 0.70 \AA$^{-1}$ along $\overbar{\mathrm{\Gamma}}\overbar{\textrm{M}}$ at 295 K and 17 K, the arrows mark the position of the three main features near E$_{F}$. (b) Sketch of the $\overbar{\mathrm{\Gamma}} \overbar{\textrm{M}}$ high-symmetry direction in the SBZ of the 1x1 surface, the SBZs of the 7$\times$7 surface along this direction are indicated. (c) ARPES intensity plot of the adatom surface bands along the $\overbar{\mathrm{\Gamma}} \overbar{\textrm{M}}$ direction at 17 K with a photon energy of 40 eV in linear color scale. (d) Negative of the second derivative of the photoemission intensity of panel (c) with respect to the energy. The circles indicate the local maxima as a function of $k$. (e) Same as (d) with the contrast increased by a factor 5 to highlight the weakest features. (f) Dispersion of the surface bands obtained by the minima of the second derivative with respect to energy. The positions of the $\overbar{\mathrm{\Gamma}}$ and  $\overbar{\textrm{M}}$ points of the 7$\times$7 SBZs are indicated.}
\end{figure}

The nature of the low-energy states of this surface can be further investigated by measuring the dispersion of the adatom bands by angle-resolved photoemission spectroscopy (ARPES). Figure \ref{fig:ARPES}(a) reports spectra of the Fermi level region measured with 40 eV photons along the $\overbar{\mathrm{\Gamma}} \overbar{\textrm{M}}_{1\times1}$ direction ($\overbar{\textrm{M}}_{1\times1}$ is the $\overbar{\textrm{M}}$ point of the unreconstructed 1$\times$1 SBZ. The sequence of the 7$\times$7 SBZ along this direction is sketched in Fig.\ref{fig:ARPES}(b)). We observe three peaks at about -0.12, -0.29 and -0.45 eV, which match the peak and the shoulders of the angle integrated spectrum of Fig. \ref{fig:PE}(a).
 The general features of our spectra are in agreement with the photoemission data of Refs. \onlinecite{uhrberg1998a, barke2006a}, except for an  energy shift caused by different treatments of the SPV effect. Our improved correction of the SPV and the high signal-to-noise ratio allow us to detect the three bands with low uncertainty in energy and in an extended momentum region (see Figs. \ref{fig:ARPES}(c)). The dispersions of these bands, marked by symbols in Figs. \ref{fig:ARPES}(d)--(f), are obtained by measuring the energies of the minima of the second derivative of the photoemission intensity with respect to the energy (Figs. \ref{fig:ARPES}(d)--(e)). These dispersions are confirmed by using the Laplacian and the curvature methods \cite{curvature} and photon energies between 24 an 75 eV (see Appendix on data reproducibility and data analysis).

\onecolumngrid

\begin{figure}[ht]
\includegraphics[scale=0.5]{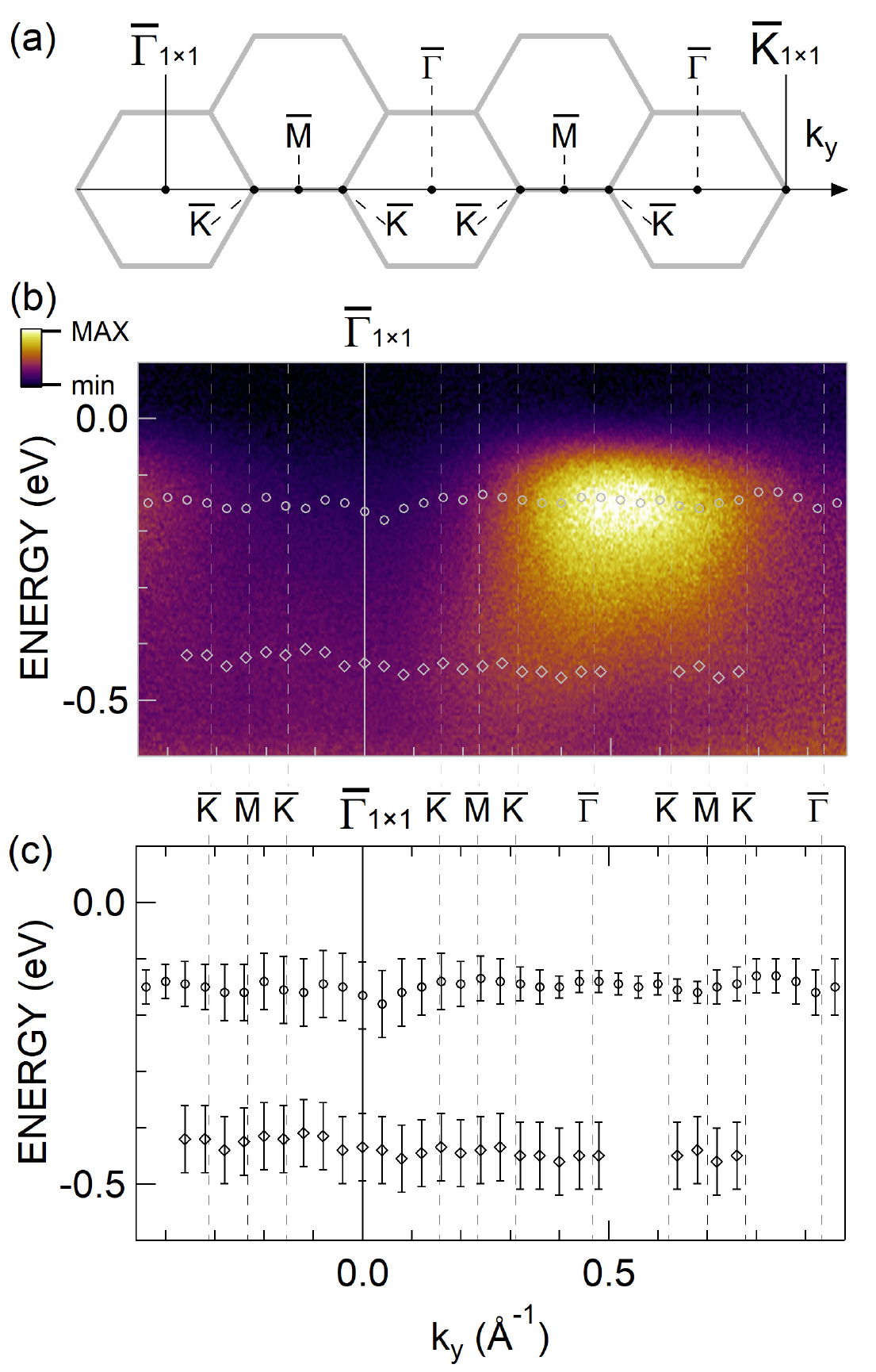}
\caption{\label{fig:ARPES-GK} (color online). (a) Sketch of the $\overbar{\mathrm{\Gamma}} \overbar{\textrm{K}}$ high-symmetry direction in the surface Brillouin zone of the ideal 1$\times$1 surface, the SBZs of 7$\times$7 surface along this direction are indicated. (b) ARPES intensity plot of the adatom surface bands along the $\overbar{\mathrm{\Gamma}} \overbar{\textrm{K}}$ direction at 17 K with a photon energy of 40 eV in linear color scale. The circles indicate the local minima of the second derivative of the intensity with respect to energy as a function of $k$. (c) Dispersion of the surface bands obtained by the minima of the second derivative with respect to energy. The positions of the $\overbar{\mathrm{\Gamma}}$,  $\overbar{\textrm{M}}$, and $\overbar{\textrm{K}}$ points of the 7$\times$7 SBZ are indicated.}
\end{figure}

Fig. \ref{fig:ARPES-GK} show the sequence of the 7$\times$7 SBZ along the $\overbar{\mathrm{\Gamma}} \overbar{\textrm{K}}_{1\times1}$ direction of the 1$\times$1 SBZ, the photoemission intensity map along this direction, and the band dispersions obtained by the second derivative method.

\begin{figure}[ht]
\includegraphics[scale=0.6]{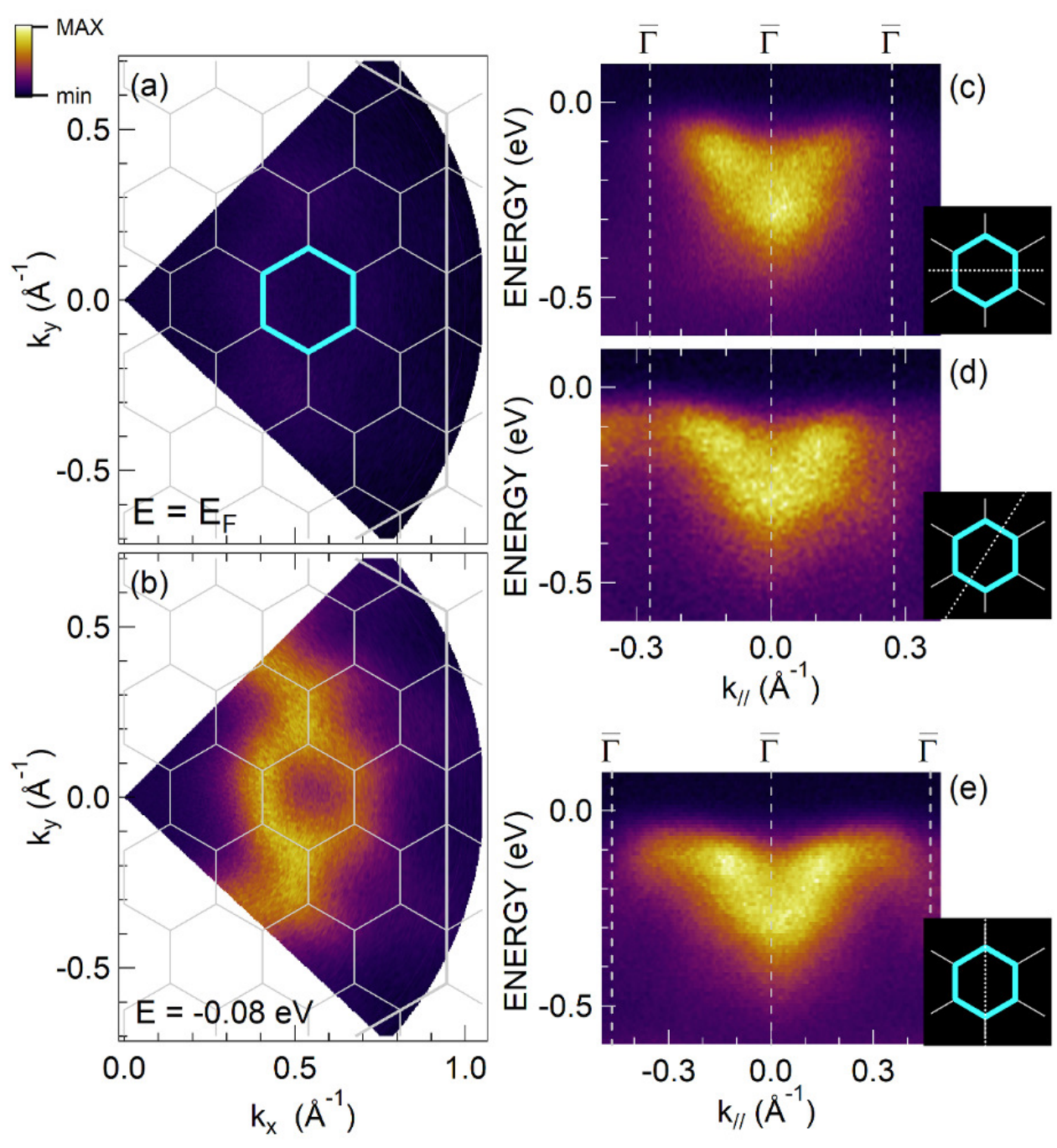}
\caption{\label{fig:ARPES-map} (color online) 
(a) Constant energy cuts of the photoemission intensity at E$_{F}$ and (b) at -0.08 eV plotted with the same color scale for a photon energy of 40 eV at 17 K. The thin white lines mark the 7$\times$7 SBZ and the thick ones are the edges of the 1$\times$1 SBZ. (c)--(e) ARPES intensity plots measured along the direction shown in small figures at the bottom-right by the dashed line in the 7$\times$7 SBZ centered at $k_x$=0.54 \AA$^{-1}$, $k_y$=0 \AA$^{-1}$ and highlighted in blue in panel (a). The wavevector scale is referred to the center of this Brillouin zone.}
\end{figure}

Figures \ref{fig:ARPES} and  \ref{fig:ARPES-GK} show that the band closest to E$_F$, centered at -0.12 eV, never crosses the Fermi level along the  $\overbar{\mathrm{\Gamma}} \overbar{\textrm{M}}_{1\times1}$ and $\overbar{\mathrm{\Gamma}} \overbar{\textrm{K}}_{1\times1}$ directions and stays at least 60 meV below E$_F$, a binding energy which is more than four times the error of the position of E$_{F}$. No other band closer to E$_{F}$ is observed for photon energies between 24 and 75 eV along high-symmetry directions. 
Additionally, we explored the electronic structure of the Si surface by ARPES over the whole 1$\times$1 SBZ. The constant energy cuts of Fig. \ref{fig:ARPES-map}(a) and (b) show that the residual intensity at E$_{F}$ derives from the tails of features observed at deeper energies broadened by the SPV effect. We also display in Fig. \ref{fig:ARPES-map}(c--e) the energy-momentum maps taken along high-symmetry directions of the 7$\times$7 SBZ marked by a thick cyan line in Fig. \ref{fig:ARPES-map}(a). In all these panels the photoemission peaks lie below E$_{F}$. 
 The low residual signal at E$_{F}$ visible at 17 K in Fig. \ref{fig:ARPES}(a) is due to the high-energy tail of the -0.12 eV band broadened by the SPV effect.
Coherently, the lack of bands across E$_{F}$ and the binding energy larger than 60 meV of the topmost band are in agreement with the deep depression of the spectral intensity at E$_{F}$ observed in our STS and angle-integrated photoemission data. Moreover, our results are also consistent with the insulator-like photoemission data reported by Uhrberg {\em et al.} \cite{uhrberg1998a}.

An intriguing feature of the band near E$_{F}$ is its dispersion. The  energy of this band varies between about -0.08 eV and -0.15 eV along $\overbar{\mathrm{\Gamma}} \overbar{M}_{1\times1}$ with alternating maxima and minima in the proximity of consecutive $\overbar{\mathrm{\Gamma}}$ points of the repeated 7$\times$7 BZ (see Figs. \ref{fig:ARPES}(d)-(f)). This behavior causes an apparent doubling of the period of the band dispersion with respect to the 7$\times$7 period. The same band appears flat within our experimental error in the $\overbar{\mathrm{\Gamma}} \overbar{\textrm{K}}_{1\times1}$ direction, which also contains $\overbar{\mathrm{\Gamma}}$ and $\overbar{\textrm{M}}$ points. 
The anomalous period of the band dispersion, the interpretation of wich will require additional investigations, beyond the scope of the present work, could be related to the presence of 
four triplets of equivalent adatoms in the unit cell (see Fig. \ref{fig:struttura}), which could give rise to modulation of the 
photoemission matrix elements in adjacent BZs by a structure factor term.
An effect of this kind was observed for the first time in graphite \cite{shirley1995a}, where the photoemission signals of the $\pi$ and $\sigma$ states  display strong intensity modulations in neighbouring BZs. The same effect in graphene causes a dispersion of the C 1s core level with a period different from that of the BZ \cite{lizzit2010a} and, in the iron-pnictide superconductor FeSe$_x$Te$_{1-x}$, distinctive signatures of reciprocal lattice symmetry breaking in the ARPES spectrum\cite{fese}.

\subsection{Magnetic phases}
Motivated by the contrasting evidences from multiple experimental measurements and first-principles DFT results on the electronic properties of the Si(111)--7$\times$7 surface, in particular by the presence of an energy gap, we extended our calculations to include possible symmetry breaking effects.
We explored the possibility of a magnetic ground state, which could be the origin of a band insulating state with an odd number of electrons per unit cell, keeping the C$_{3v}$ point-group symmetry of the surface in the DAS model. Starting from the non-magnetic band structure in Fig. \ref{fig:bands_normal}, the exchange splitting of a ferromagnetic surface should be at least 0.35 eV to shift the first minority adatom band high enough in energy to cause the complete filling of the five lowest majority adatom bands with the five adatom electrons and an energy gap at E$_{F}$ between the fifth majority band and the others. A smaller exchange splitting could still give rise to a magnetic insulator in the case of a breaking of the C$_{3v}$ symmetry that removes the degeneracy of the  adatom bands at $\overbar{\mathrm{\Gamma}}$ at 0.02 eV and 0.04 eV.

\begin{figure}[ht]
\includegraphics[trim=1.1cm 0 0 -0.cm, scale=0.55]{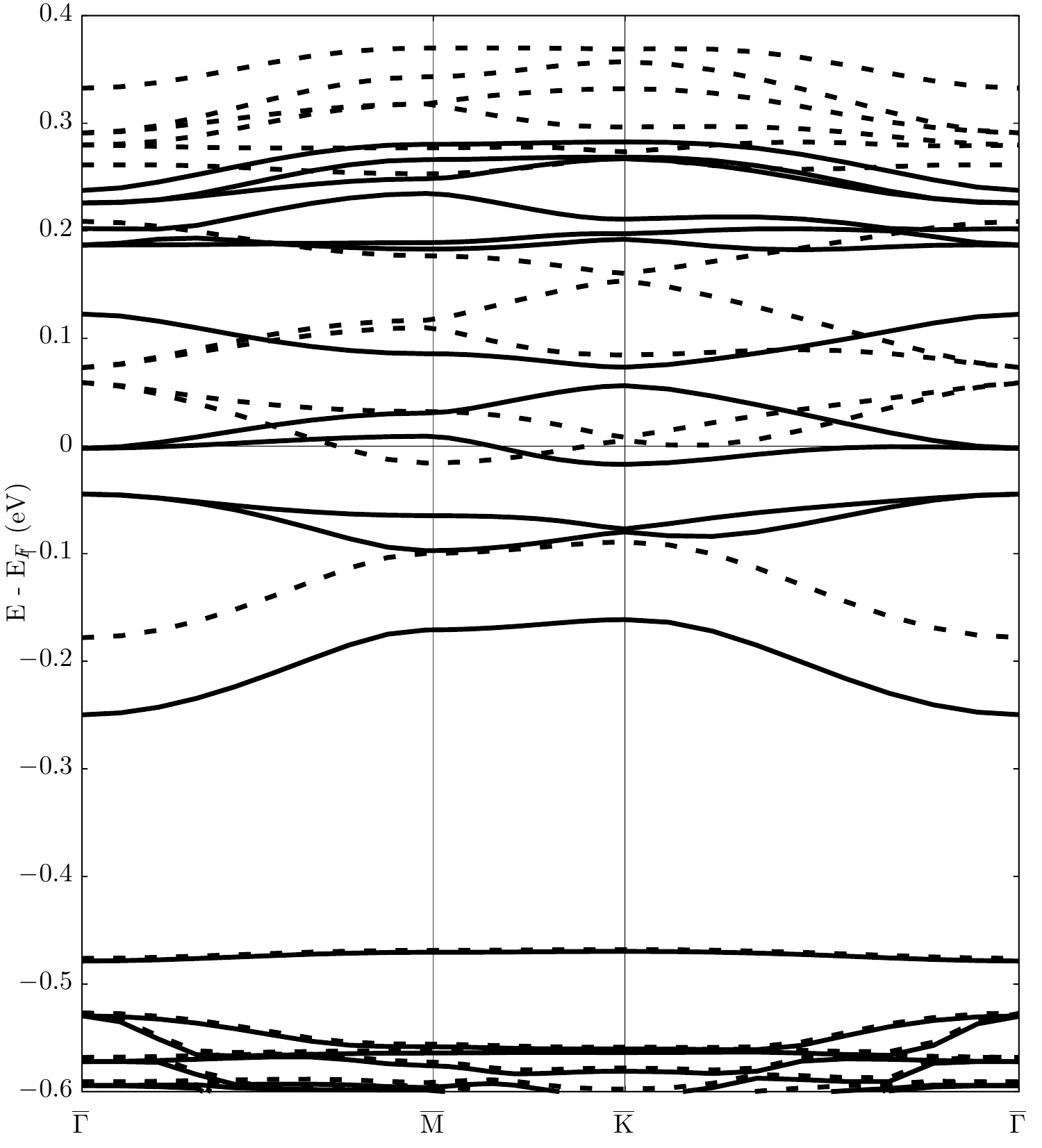}
\includegraphics[scale=0.65]{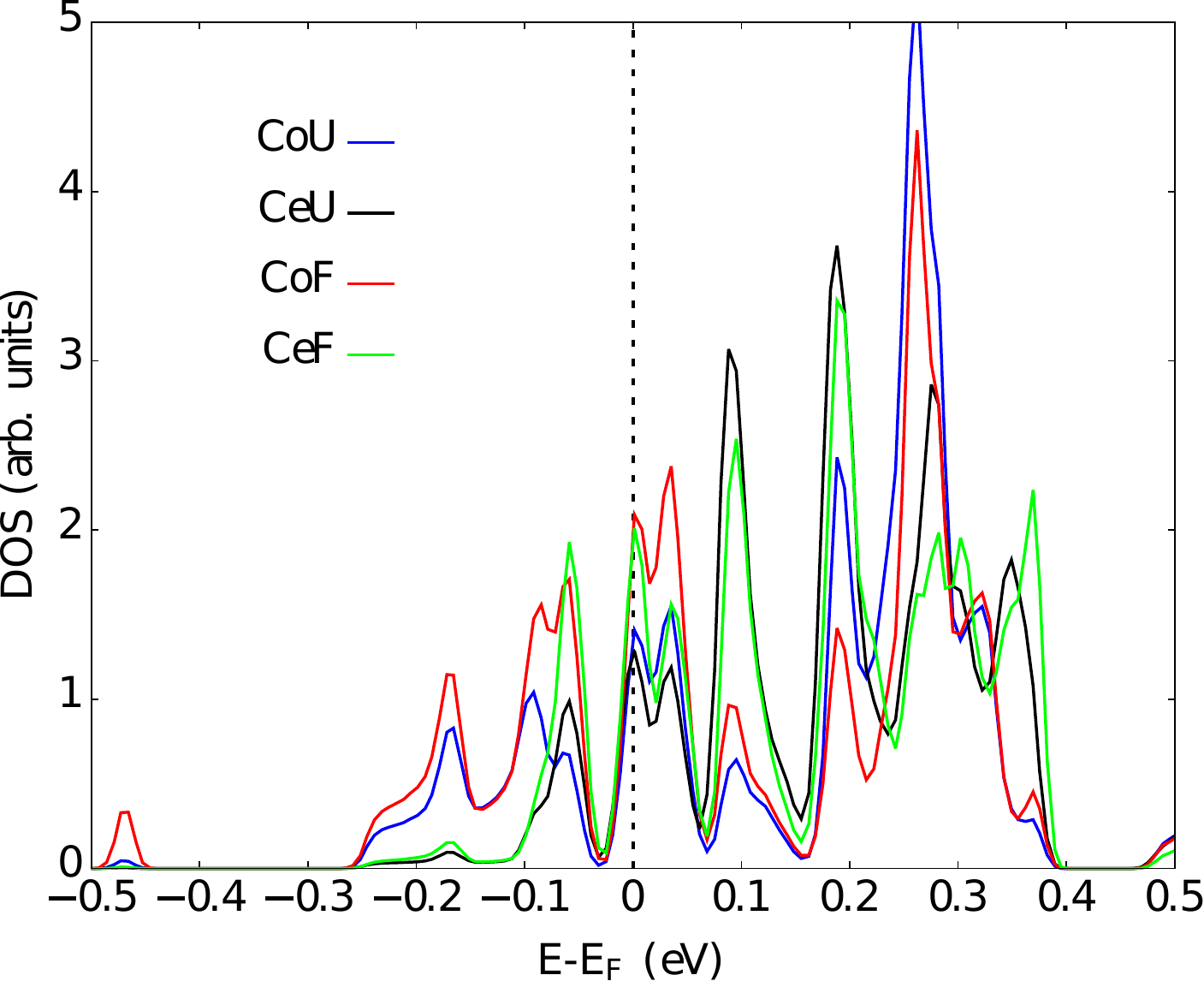}
\caption{(Upper panel): Band structure of the ferromagnetic Si(111)--7$\times$7 surface. Solid (dashed) lines represent the majority (minority) bands. (Lower panel): (color online) Total (majority+minority) PDOS on adatom sites for the ferromagnetic Si(111)--7$\times$7 surface.
Different colors represent  corner (blue) and central (black) adatoms in the unfaulted region and  corner (red) and central (green) adatoms in the faulted region. }
\label{fig:ferromagnetic-bands}
\end{figure}

We tested this possibility allowing a spin polarization of the surface bands. The theoretical stability of a possible magnetic phase of this surface was not clearly solved in the literature, which contains conflicting results\cite{sheka2003, mag-silicon}.
We performed a self-consistent calculation which starts from parallel non-zero magnetic moments on Si adatoms and converges to a ferromagnetic solution with a total magnetization of 2.3 $\mu_B$ per unit cell.
The total energy of the magnetic solution, within our approximations, is about 0.5 meV/adatom lower than the non-magnetic solution, indicating nearly degenerate phases.
The band structure of the ferromagnetic phase, reported in Fig. \ref{fig:ferromagnetic-bands}, shows an almost rigid shift between the majority and minority adatom bands with an exchange splitting of about 0.1 eV, which is less than the 0.35 eV necessary for an insulating state.

The band structure within 0.3 eV below E$_F$ has three (one) completely filled majority (minority) bands. Partially filled spin-up and -down bands eventually lead to a metallic ground-state, still in disagreement with the experiments.
It is interesting, however, to calculate the effect of exchange splitting on the PDOS on the adatom sites, presented in Fig. \ref{fig:ferromagnetic-bands}, summing the majority and minority contribution in order to compare with non-spin-polarized STS results of Fig. \ref{fig:sts}.
The DOS at E$_F$ is reduced by $\sim$ 30\% with respect to the non-magnetic PDOS and the centroid of the occupied PDOS is shifted by about 50 meV towards higher binding energies with respect to the non-magnetic solution, thus partially improving the agreement with the experimental results. As evident from Fig. \ref{fig:ferromagnetic-bands} the occupied manifold formed by four bands (three of majority spin and one of minority) is well  separated from the bands at E$_{F}$ by an energy gap of $\sim$ 0.05 eV accommodating four of the five adatoms dangling bond electrons. The remaining electron is now responsible for the metallic nature of the surface because of the overlap of majority and minority bands at E$_F$.

\begin{figure}[h]
\includegraphics[scale=0.45]{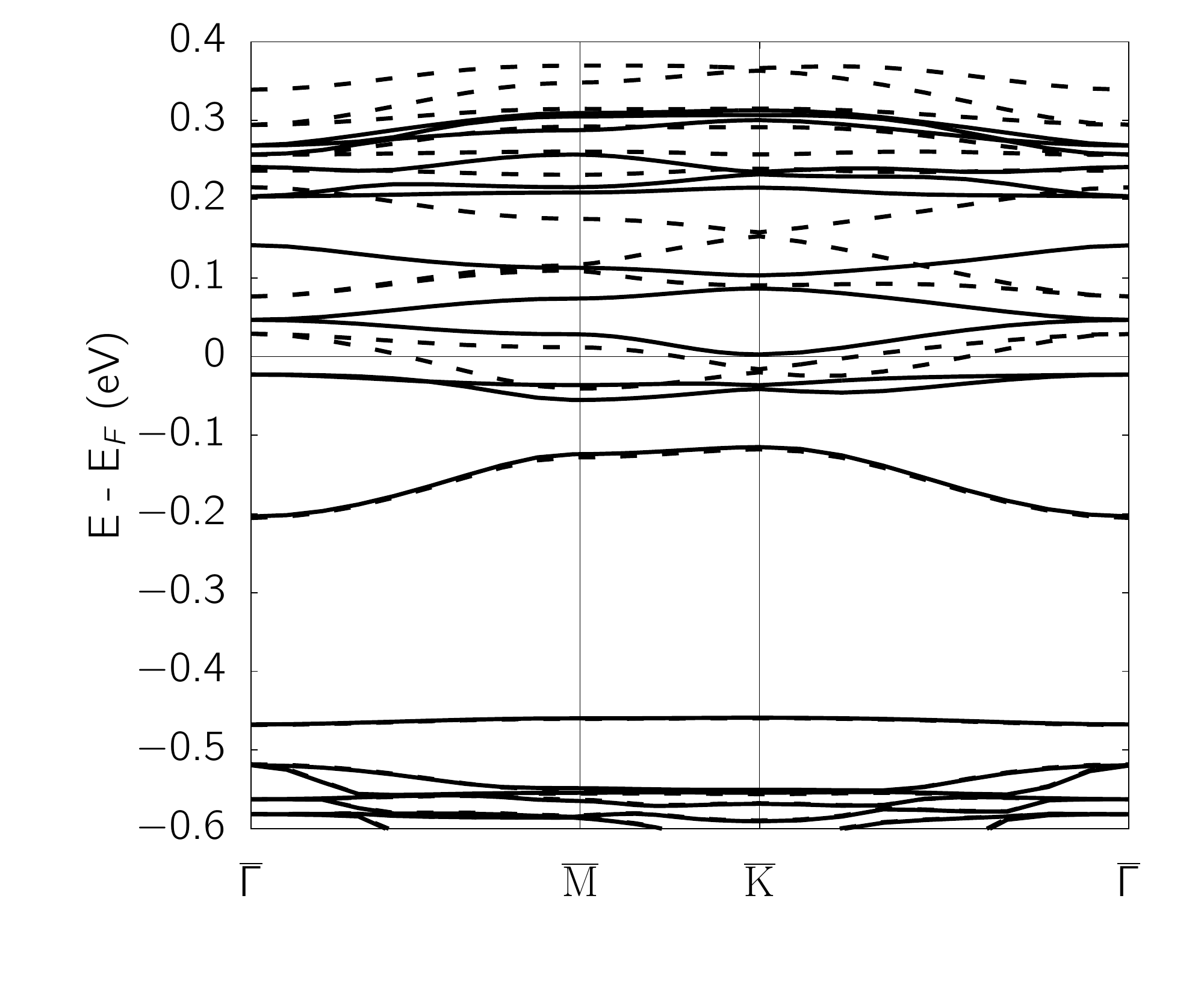}
\caption{\label{fig:ferrimagnetic-bands} Band structure of the ferrimagnetic Si(111)-- 7$\times$7 surface. Solid (dashed) lines represent the majority (minority) bands.}
\end{figure}

The magnetic moment of the surface is mainly localized on the adatoms of the faulted region, followed by the central adatoms of the unfaulted region and with a marginal contributions from corner adatoms of the unfaulted half.
We also tested two other starting spin configurations including antiparallel guesses for the Si adatom spin polarizations.
In particular, we considered $(i)$ a starting configuration in which the corner adatoms in the same (faulted or unfaulted) region are ferromagnetically aligned to each other while the two regions are  antiferromagnetically coupled and  $(ii)$ the corresponding one for the central adatoms. 
In both cases the solution converges to the same metallic ferrimagnetic phase with a net magnetic moment of  $\simeq$ 1.0 $\mu_B$ and an energy about 1 meV/adatom lower than that of the non-magnetic case. The relative band structure is shown in Fig. \ref{fig:ferrimagnetic-bands}.  
Interestingly, the system is half-metallic, showing a small insulating gap at E$_{F}$ in the majority spin channel, while remaining metallic in the minority one. That is because the exchange splitting, like in the ferromagnetic case, is not larger than 0.1 eV, not enough to avoid that E$_{F}$ intersects a manifold of bands degenerate at $\overbar{\mathrm{\Gamma}}$.

It is conceivable that a breaking of the C$_{3v}$ symmetry, caused by a structural or charge distortion that removes the degeneracy at $\overbar{\mathrm{\Gamma}}$, might induce an insulating state. That would represent a sort of band Jahn-Teller effect -- either static or dynamic. For instance, in the case of the ferromagnetic solution, a structural distortion-induced splitting might separate the fourth majority adatom band from the others and move it below E$_{F}$. In this way the system could have bands with integer occupation and become a magnetic insulator.
We investigated if the inclusion of electronic correlation effects at the level of the DFT+U approximation\cite{LDA+U} can increase the exchange splitting near the Fermi level and/or stabilize a structural distortion that lowers the symmetry and make the surface insulating. We used a starting ferromagnetic configuration with random vertical displacements of the adatoms (with a maximum displacement of 0.2 \AA). Structural relaxation in the DFT+U approximation, with the Hubbard repulsion term U set to a reasonable value of 2.0 eV\cite{profeta2007a} and the exchange integral J set to 0.8 eV, fails to find an insulating state because the exchange splitting remains of the order of 0.1 eV near the Fermi level with no indication of a structural distortion. Indeed, the obtained band structure closely resembles the ferromagnetic solution shown in Fig. \ref{fig:ferromagnetic-bands}.
This result does not in itself exclude some kind of correlation-assisted Jahn-Teller distortion. This type of possibility was met before, for example in $K_4C_{60}$, a narrow-gap Jahn Teller distorted insulator,~\cite{kiefl}  which DFT+U calculations fail to describe, always stabilizing an undistorted metal.\cite{capone}. While we will return to this point in the discussion below, it is necessary to conclude here that no standard first-principles, mean-field approach we attempted succeeds to explain the experimental insulating state of Si(111)--7$\times$7.

\section{ Discussion and Conclusions} 
\label{sect:discussion} 

Our complementary STS and photoemission data on the Si(111)--7$\times$7 surface agree showing a strongly depleted or vanishing density of states at E$_{F}$ at and below 17 K, independently of bulk doping, and no surface band crossing the Fermi level. The binding energy of the band closest to E$_F$ measured by ARPES is larger than 60 meV, confirming the presence of an energy gap and a genuine insulating state. The small residual photoemission intensity at the nominal energy of the Fermi level in our 17 K spectra is explained by the 50 meV SPV-induced extrinsic broadening of the spectra, comparable to the binding energy of the topmost filled surface band. 
The proper treatment of SPV shift is crucial to resolve inconsistencies among previous photoemission studies\cite{Demuth1983a, uhrberg1998a, losio2000a, barke2006a}, which report conflicting results.
In addition, the differential conductance data presented in Section \ref{sec:sts} indicate that the electronic properties of the three CeU adatoms are appreciably more affected by the temperature than those of the other nine adatoms in the unit cell.

These experimental evidences are in contrast to the computational results of Section \ref{theory} which predict a metallic surface with a peak of the DOS at E$_{F}$ and without relevant differences in the PDOS(E$_F$) among the adatoms (not larger than a factor 2.5), in agreement with previous calculations\cite{fujita1991a, ortega1998a, brommer1993a, smeu2012a}. 
This suggests that the proper description of the electronic properties of Si(111)-- 7$\times$7 surface should require additional ingredients with respect to the present calculations. 
A first step along this hypothesis was done performing a spin-polarized DFT calculation, finding different marginal magnetic instabilities of the surface which give rise to a finite magnetic moment.
The band structure and density of states near the Fermi energy of the computed ferromagnetic and ferrimagnetic solutions are modified with an average shift of the occupied adatom states to higher binding energies with respect to the nonmagnetic bands. However, the studied magnetic solutions remain metallic because overlapping adatom bands degenerate at $\overbar{\mathrm{\Gamma}}$ are still partially filled, due to the small exchange splitting.

We have further explored the possibility of a larger exchange splitting and a structural symmetry breaking taking into account strong correlation effects, within DFT+U approximation, without finding indications of an insulating state or structural distortions. However, local distortions, if not intrinsic of the perfect surface, could be still triggered by a long range effect of surface defects or subsurface dopant impurities, which are not considered in the present model.

Other missing elements that could also explain discrepancies between  DFT predictions and the experimental results are correlation effects beyond the DFT+U level and strong electron-phonon coupling.
Electron correlation effects, caused by both intra- and inter-site Coulomb  repulsion, are not properly described within our mean-field DFT and DFT+U calculations and could lead to a Mott-Hubbard insulator or Wigner-like 2D crystallization.
The presence of significant electron correlations is indeed consistent with the short lifetimes of the electronic surface state studied by ultrafast spectroscopy \cite{mauerer2006a} and with the value of the screened on-site Coulomb interaction on the Si adatoms estimated to be 1 eV \cite{ortega1998a, hansmann2013b}, twice larger than the energy range of the dangling-bond states of the adatoms.

We note that to eventually employ a better theoretical approach than DFT+U, for example DFT+ dynamic mean field theory (DMFT) ~\cite{anisimov1997},
a calculation presently impossible for Si(111)--7$\times$7 because of its inordinate size, one should begin with the ferrimagnetic starting point.  Antiferromagnetic spin waves gives rise to stronger quantum fluctuations than ferromagnetic ones, and a distortion-induced gap may ensue to stabilize the system. 

On the other hand, electron-phonon interaction might lead, especially in conjunction with carrier localization, to some form of electron self-trapping \cite{toyozawa1961a,toyozawa1980a}. In simple words, the electron's negative charge may distort the ion lattice around itself, so much to give rise to a localized attractive potential that binds and traps the electron itself, giving rise to a self-trapped polaron with large effective mass, further enhancing electron correlations. 
The electron-phonon coupling constant measured by Barke {\em et al.} \cite{barke2006a}
is $\lambda \sim$1, which could be large enough to induce such effects.
Although speculative, this tentative scenario could nonetheless explain our observations as well as those of Refs. \onlinecite{persson1984a,schillinger2005,dangelo2009a}.
This scenario appears to be supported by the STS temperature dependence.  Specifically, the filled electronic states nearest to the E$_F$, which  in the high temperature metallic state are relatively evenly spread over all the unit cell adatoms, localize mostly onto the three near-vacancy corner faulted CoF  adatoms  in the low-temperature insulating state.
Experimentally, recent applications of non-contact pendulum-AFM  spectroscopy, ~\cite{stipe???} an ultra-sensitive technique that can detect and map surface electronic excitations and their changes, suggest that more precise information on the nature of this metal-insulator transition should be derived, similar to those obtained for the normal-superconductor transition~\cite{kisiel2011}
or the valence changes of surface oxygen vacancies in SrTiO$_3$~\cite{kisiel2018}.

In summary, our experimental and theoretical results call for a critical revision of the low-temperature electronic properties of the Si(111)--7$\times$7 surface, which is not merely a standard test sample, but poses difficult problems undergoing interesting phenomena and instabilities, still in need to be further clarified.

\begin{acknowledgments}
We are grateful to G. Giovannetti, A. Baraldi and M. Fabrizio for discussions. P. M., P. M. S., and C. C. acknowledge the “Progetto Premiale, Materiali e Dispositivi  Magnetici e Superconduttivi per Sensoristica e ICT” and the project EUROFEL-ROADMAP ESFRI of the Italian Ministry of Education,  University and Research (MIUR). E. T. acknowledges partial support by European Research Council Advanced Grant N. 834402 ULTRADISS,
and by the Italian Ministry of University and Research through PRIN UTFROM N. 20178PZCB5.
G. P. acknowledges support from CINECA Supercomputing Center through the ISCRA project and financial support from the Italian Ministry for Research and Education through the PRIN-2017 project “Tuning and understanding Quantum phases in 2D materials - Quantum 2D (IT-MIUR Grant No. 2017Z8TS5B)
\end{acknowledgments}

\appendix*

\section{Surface photovoltage correction}

The photoemission spectra of semiconductor surfaces are affected by energy shifts and broadening caused by the SPV effect \cite{alonso1990a, marsi1998a} and by the charge carrier transport between the sample holder and the surface hit by the photons, which induce an uncertainty in the position of the Fermi level in the spectra. 
\begin{figure}[ht]
\includegraphics[scale=0.3]{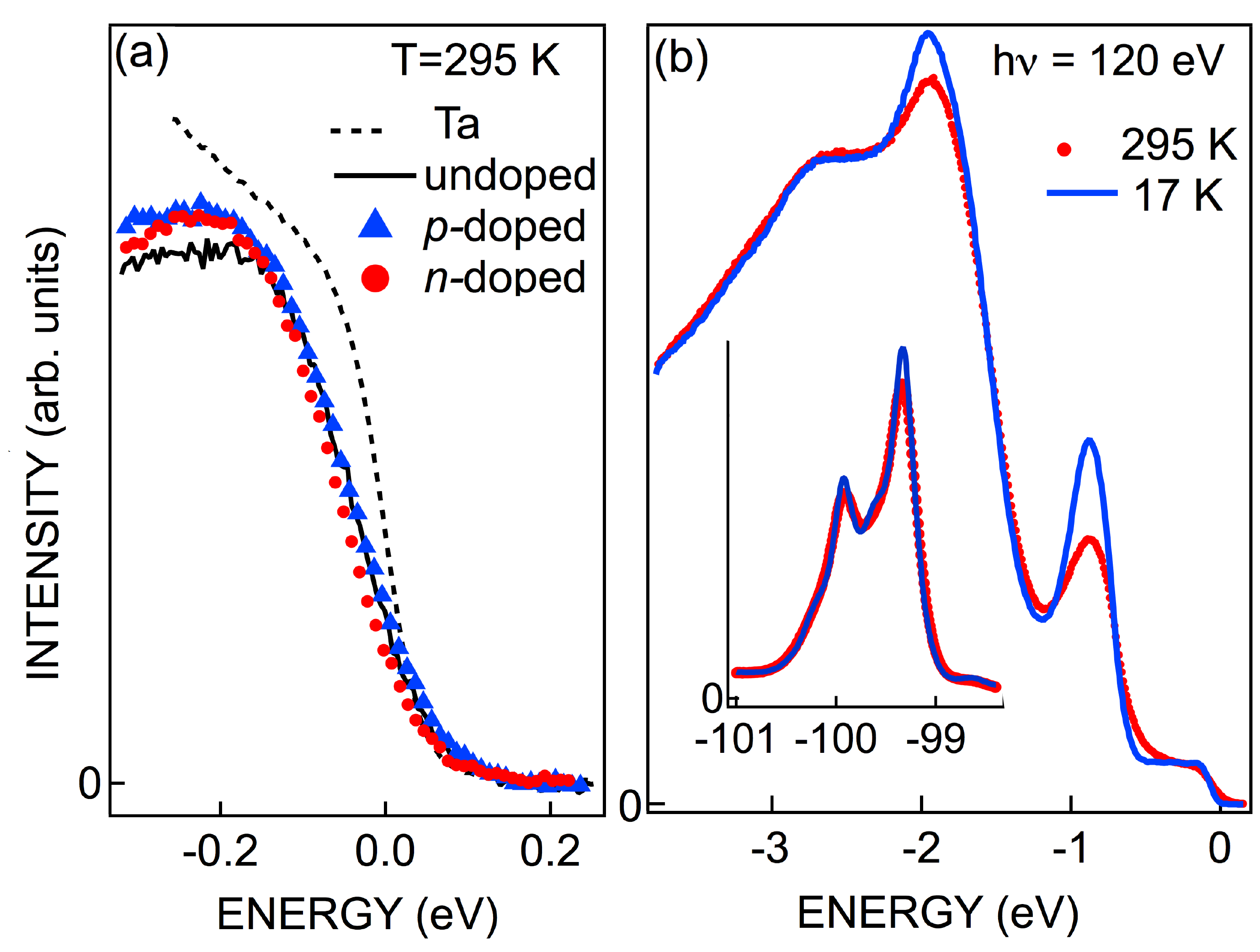}
\caption{\label{fig:SPV}
(color online). (a) Integrated photoemission spectra of $p$-doped, $n$-doped and undoped samples at 295 K compared to the spectrum of Ta. The shift between the spectra of the $p$-doped and $n$-doped samples, caused by the surface photovoltage, provides the maximum error on E$_{F}$ at 295 K. (b) Spectra of the surface states and valence band region, and of the Si 2p core levels (inset) at 295 K and 17 K measured with 120 eV photons on a $p$-doped sample. The 17 K spectra have been rigidly translated by -414 meV to correct for the surface photovoltage shift at low temperature. This shift has been calculated by minimizing the differences between the spectra at the two temperatures in the core level region and in the region of the valence band and surface states below -0.5 eV. The E$_{F}$ region of these spectra is reported in Fig. \ref{fig:PE}(b).}
\end{figure}
Its location and maximum error at 295 K in our data were determined by comparing the spectra taken on 7$\times$7 surfaces of $n$-, $p$-doped and undoped samples, and on an Ag film deposited on Si and surrounded by the clean 7$\times$7 surface, with the spectrum of a Ta plate on the sample holder (see Fig. \ref{fig:SPV}(a)). The opposite SPV shifts of the spectra of $n$-doped and $p$-doped samples cause a maximum difference of 15 meV between the positions of high-energy cutoffs of the spectra, in agreement with Ref. \onlinecite{barke2006a}, which corresponds to a maximum error of  ±8 meV in the position of E$_{F}$ at 295 K. To determine the SPV and carrier-transport shift at 17 K we measured the shifts of the Si 2p core level peaks, of the restatoms peak at -0.9 eV, and of the sharp features of the Si valence band cooling from 295 K to 17 K while keeping a constant photon flux at 120 eV.  These shifts are the same within the experimental maximum error of ±5 meV for the Si 2p core levels and the -0.9 eV peak, and of ±20 meV for the valence band features. Fig. \ref{fig:SPV}(b) shows that a rigid shift of -414 meV of the 17 K spectrum minimizes the differences between this spectrum and the RT spectrum both in the core level and in the valence band region in a $p$-doped sample. 

Therefore, we assume that the best estimate of the additional SPV shift in our low-temperature spectra is given by the rigid shift measured using the above-mentioned features at constant photon flux, while we use the shift of the valence band and of the -0.9 eV peak for photon energies lower than 120 eV. An additional source of error is a possible random shift less than 3 meV of the photon energy during the spectra acquisition. We estimate that the maximum error of the position of E$_{F}$ at 17 K is thus $\pm$ 15 meV by adding the error of the SPV at 295 K, that of the rigid shift  from room to low temperature, and the energy stability of the photon beam.

\appendix*

\section{Data reproducibility, data analysis, and effects of defects}

\begin{figure}[ht]
\includegraphics[scale=0.3]{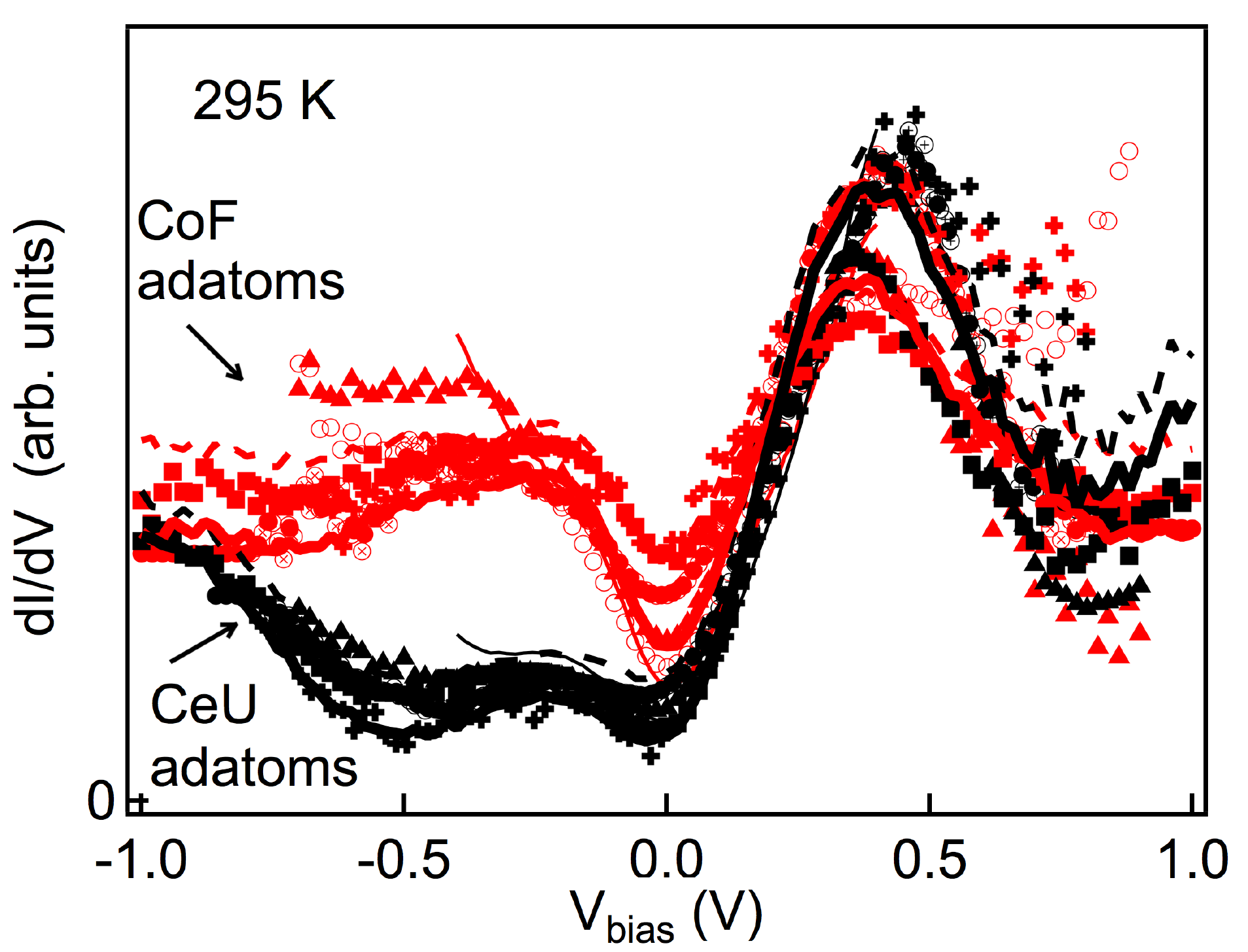}
\caption{\label{fig:reprod_RT}
(color online). Tunneling spectra measured at 295 K with Au tips with different structures on CoU and CeF adatoms. Pairs of spectra measured with the same tip one after the other are drawn with the same type of line or symbol.}
\end{figure}

\begin{figure}[ht]
\includegraphics[scale=0.3]{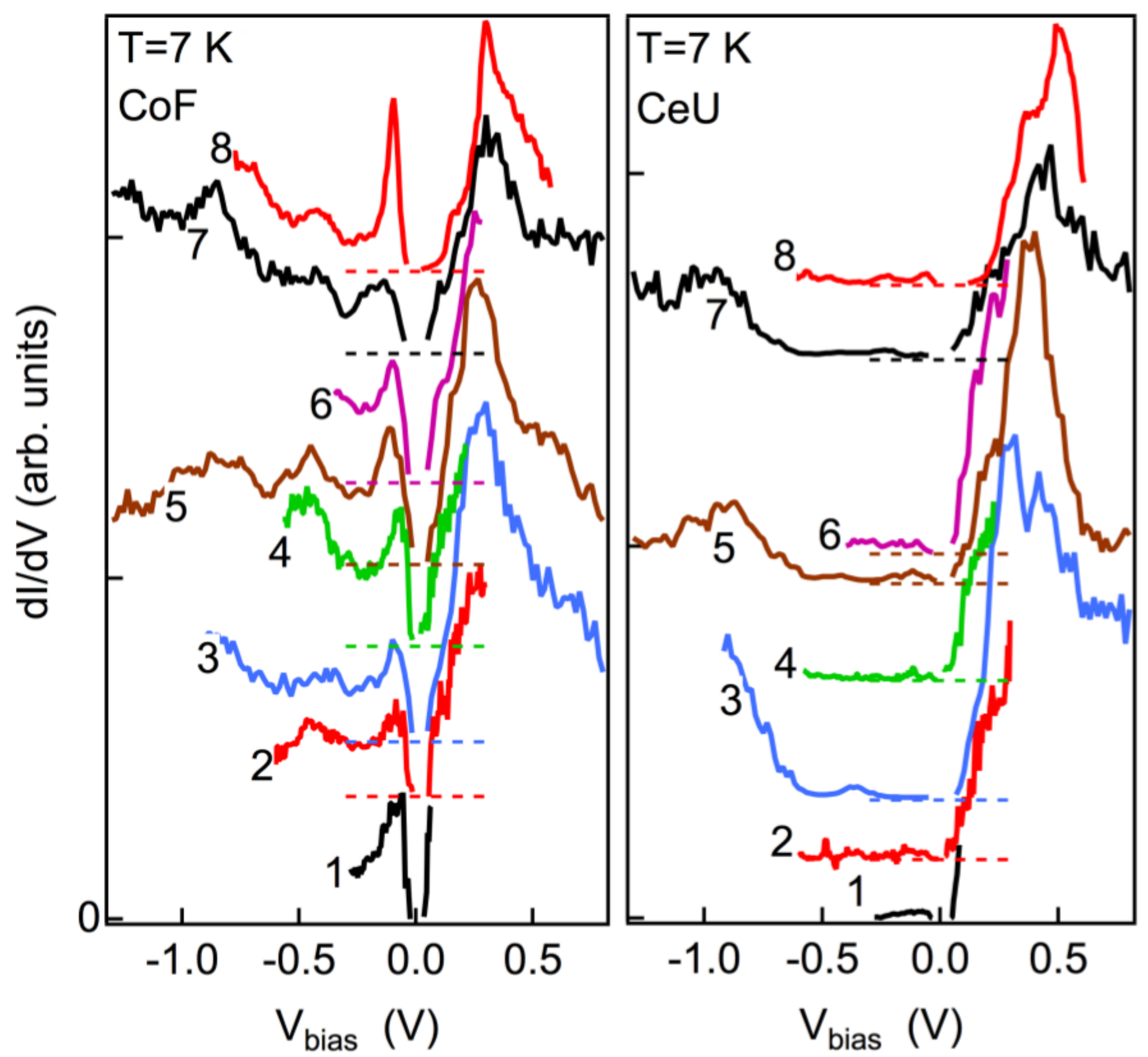}
\caption{\label{fig:reprod_7K}
(color online). Tunneling spectra measured at 7 K with Au tips with different structures on CoF and CeU adatoms. Spectra measured with the same tip one after the other are tagged by the same number. The spectra are shifted upwards for clarity, the dashed lines indicate the zero level for each spectrum.}
\end{figure}

The variability of the tunneling spectra acquired with different Au tips on the CoF and CeU adatoms is shown in Figs. \ref{fig:reprod_RT}  and \ref{fig:reprod_7K} for 295 K and 7 K respectively. Spectra of the two adatoms plotted with the same symbol in Fig. \ref{fig:reprod_RT} or tagged with the same number in Fig. \ref{fig:reprod_7K} were acquired with the same tip at the same time.  
\begin{figure}[ht]
\includegraphics[scale=0.6]{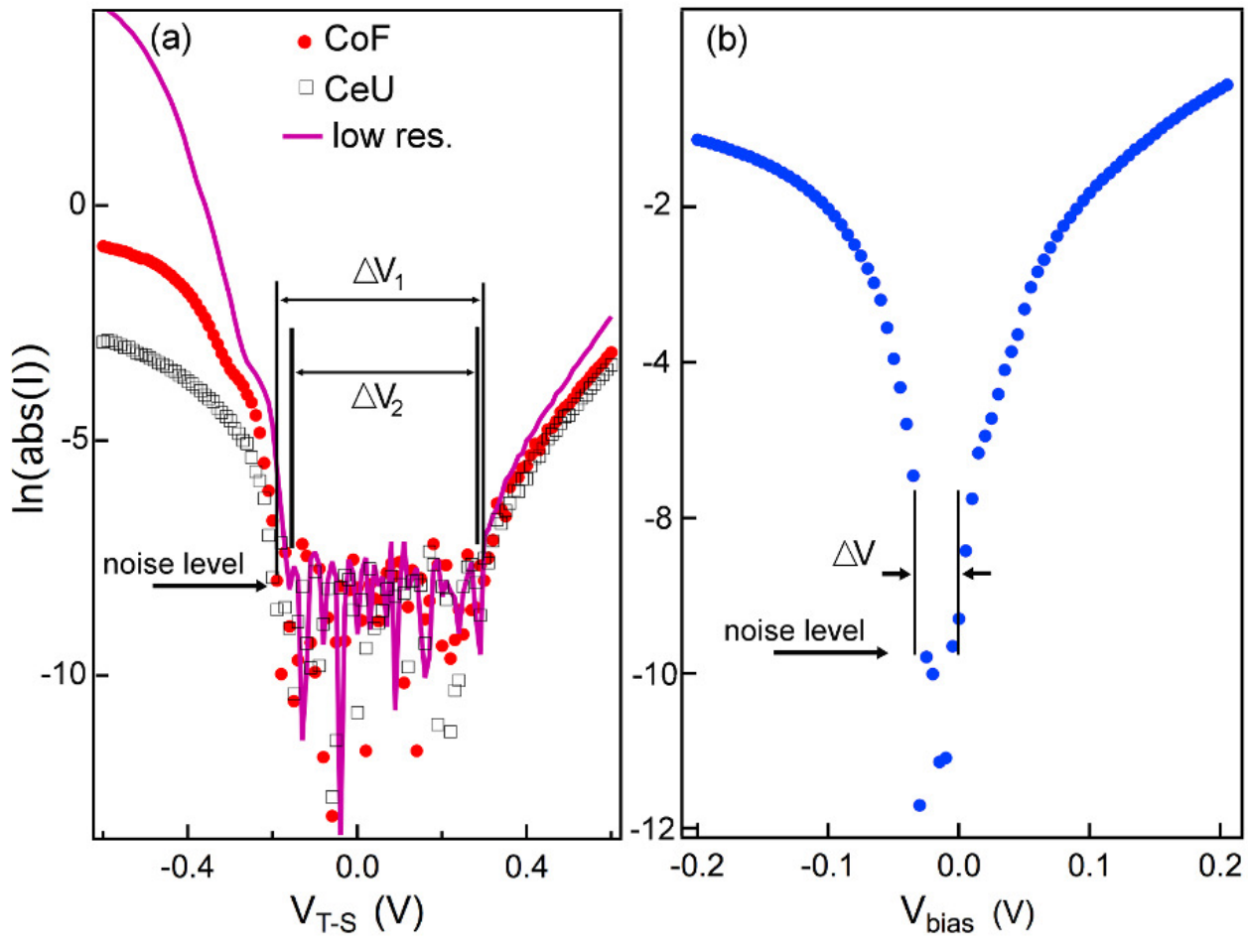}
\caption{\label{fig:gap}
(color online). (a) Natural logarithm of the tunneling current I measured on the CoF and CeU adatoms at high tunneling resistance (set point I$_{tunneling}$=1 pA, V$_{T-S}$=3 V) and at low tunneling resistance (set point I$_{tunneling}$=400 pA, V$_{T-S}$ =3 V) at 7 K plotted as a function of the voltage bias between the sample holder and the tip V$_{T-S}$. At low tunneling resistance the curves of the two adatoms are similar, since the transport between the surface and the sample holder dominate the tunneling contribution. The difference between the voltage gaps measured in the two conditions gives the surface gap. The data refer to a single acquisition cycle. (b) Natural logarithm of the tunneling current as a function of the voltage bias between the surface below the tip and the tip. The current in this panel is the sum of the contributions of all the adatoms.}
\end{figure}
\begin{figure}[ht]
\includegraphics[scale=0.5]{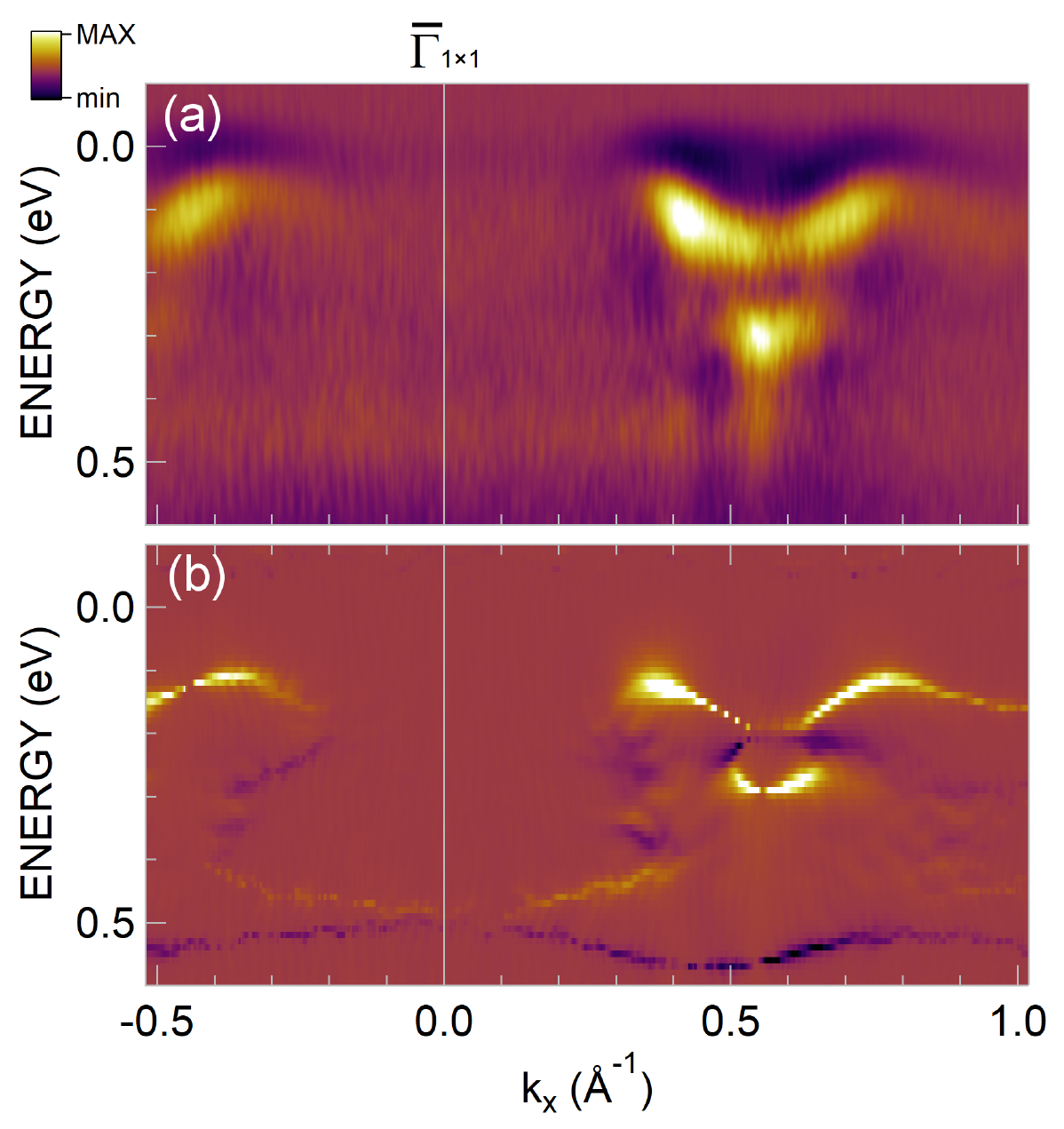}
\caption{\label{fig:Lap_Curv}
(color online). Laplacian of the photoemission intensity data of Fig.\ref{fig:ARPES}(c), the noise derives mainly from the derivative with respect to k. (b) curvature plot of the intensity data of Fig.\ref{fig:ARPES}(c)}
\end{figure}
\begin{figure}[ht]
\includegraphics[scale=0.6]{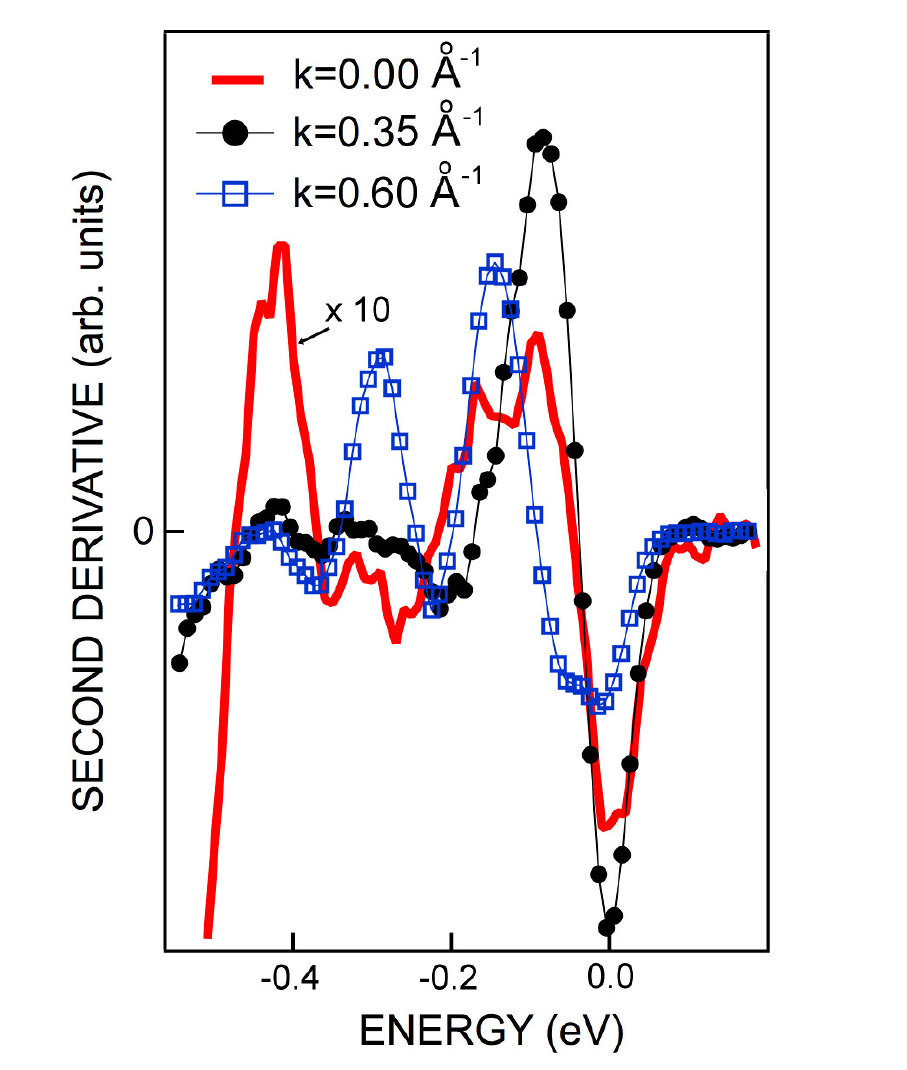}
\caption{\label{fig:Gamma}
(color online). Negative of the second derivative of the photoemission intensity of Fig. \ref{fig:ARPES}(c) with respect to energy at the wavevectors indicated in the figure. The curve for k=0 has been multiplied by a factor ten. The maximum near -0.12 eV is particularly broad at k=0 \AA\ $^{-1}$}
\end{figure}
\begin{figure}[ht]
\includegraphics[scale=0.3]{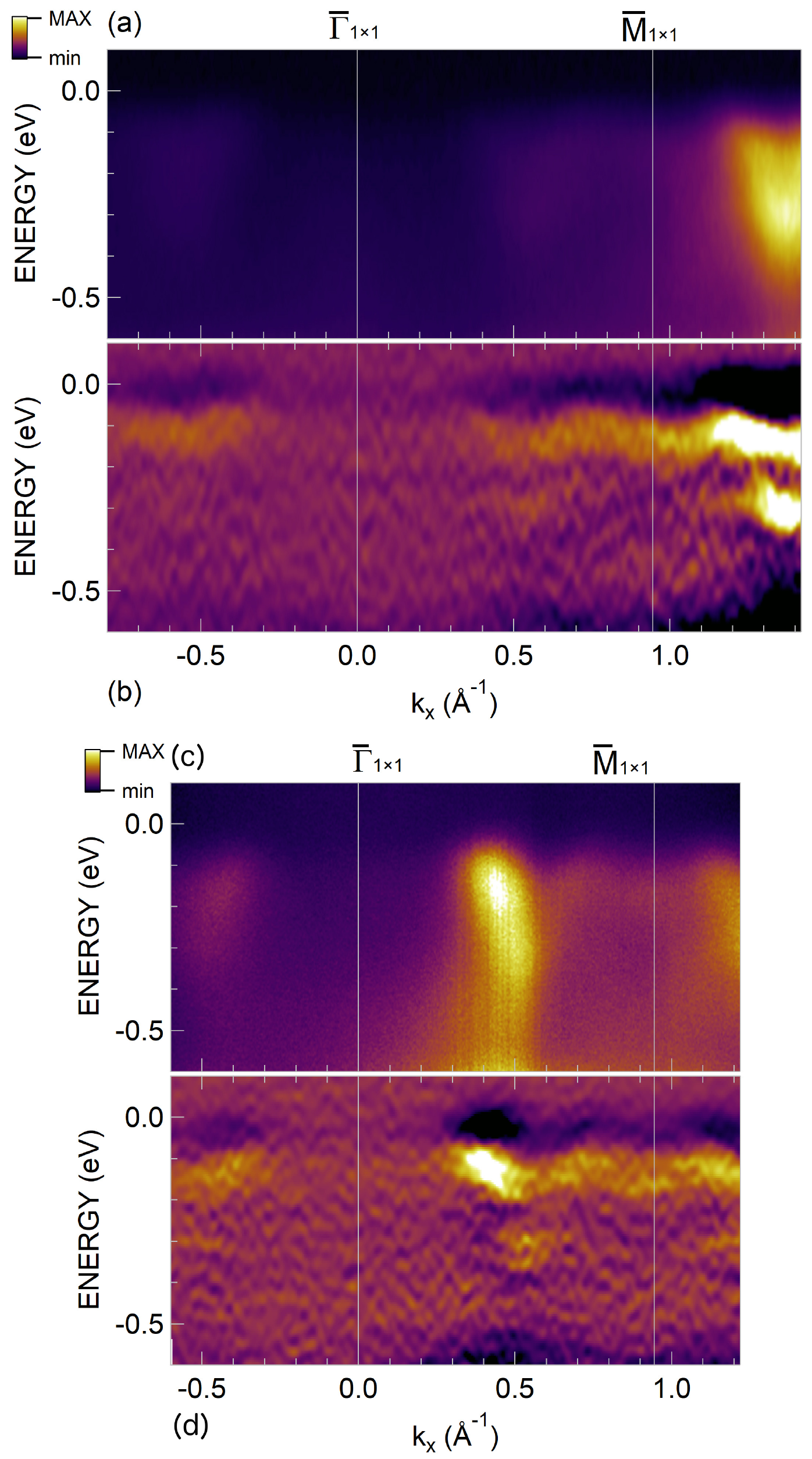}
\caption{\label{fig:AltreEnergie}
(color online). (a) ARPES intensity plot along the $\overbar{\mathrm{\Gamma}}\overbar{\textrm{M}}$  direction at 17 K with a photon energy of 75 eV. (b) Negative of the second derivative of the photoemission intensity of panel (a) with respect to energy. (c) and (d) Same as (a) and (b) with a photon energy of 60 eV }
\end{figure}
\begin{figure}[ht]
\includegraphics[scale=0.6]{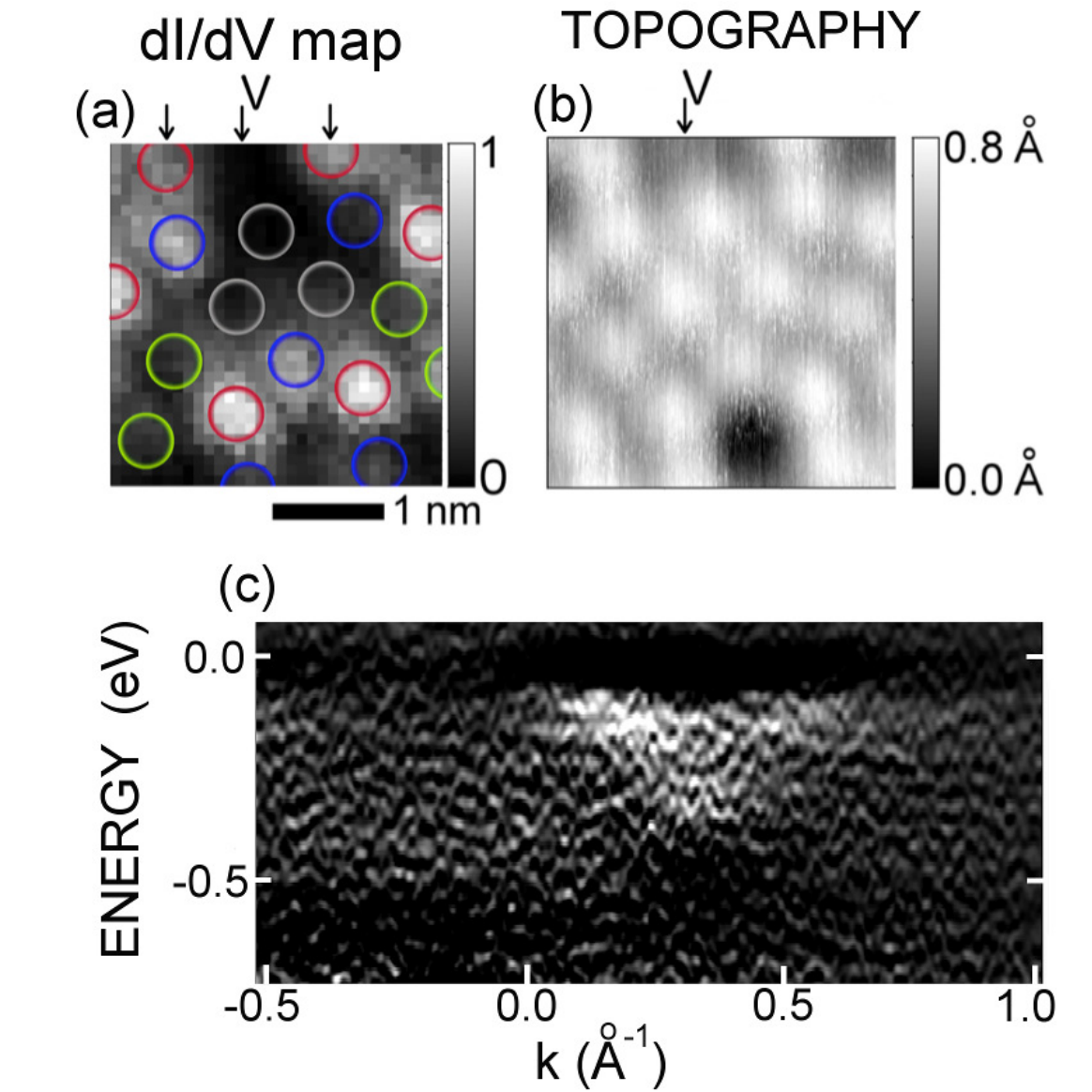}
\caption{\label{fig:Difetto}
(color online). (a) dI/dV map between -0.2 V and -0.1 V of the region with a defect shown by the topographic map in (b). The set point was V$_{bias}$=1.4 V and I$_{tunneling}$=5 pA. V indicates the missing adatom.  The red, blue, green and gray circles mark the CoF, CoU, CeF and CeU adatoms respectively. (c) Negative of the second derivative with respect to energy of the ARPES intensity plot along the $\overbar{\mathrm{\Gamma}}\overbar{\textrm{M}}$ direction of a surface with high defect density at 17 K.}
\end{figure}
Each pair of spectra was taken with a different tip.  Spectra obtained with insulating or semi-insulating tips (with low intensity near E$_F$) are not reported and were not used for the analysis. The difference in the spectral intensity between -0.2 and 0.0 V between CoF and CeU adatoms is always observed, and reaches a factor of at least ten in all the cases at 7 K.

The peaks at -0.10 V and  -0.45 V are always observed on the CeU adatoms at low temperature.  The differences between the spectra obtained with different tips are mainly in the relative intensities of the parts at positive and negative voltages and in the intensity of the background at high bias voltages.

Fig. 15(a) shows the logarithm of the tunneling current as a function of the voltage bias between the sample holder and the tip (V$_{T-S}$) measured on the CoF and CeU adatoms at high and low tunneling resistance. The voltage gap $\mathrm{\Delta}$V$_{2}$ in the low tunneling resistance curve is mainly caused by the transport of charge carriers between the surface below the tip and the sample holder and includes possible Coulomb blockade effects. The difference between the similar gap $\mathrm{\Delta}$V$_{1}$ in high resistance curves and $\mathrm{\Delta}$V$_{2}$ is the surface gap [21]. The same gap $\mathrm{\Delta}$V is also visible when the logarithm of the current is plotted vs the voltage bias between the tip and the surface (Fig. 15(b)).

The analysis of the ARPES data of Fig. \ref{fig:ARPES}(c) with the Laplacian method or with the curvature method, show in Fig. \ref{fig:Lap_Curv}, provides the same dispersion of the surface bands obtained with the second derivative with respect to energy reported in Fig. \ref{fig:ARPES} within the experimental error. The increased noise level in these plots derives mainly from the derivative with respect to the wavevector k.
The energy of the topmost band near $\overbar{\mathrm{\Gamma}}$ of the 1$\times$1 SBZ is reported with a large error in Fig. \ref{fig:ARPES}(d) and (f) because the maximum of the negative of the second derivative with respect to the energy is particularly broad for this wave vector, as shown in Fig. \ref{fig:Gamma}. The width of the peak of the second derivative at k=0 \AA\ $^{-1}$ is about twice those at larger wave vectors. A possible interpretation is the presence of two components, one at about -0.08 eV and the other at -0.14 eV. These two components and the structure factor mechanism mentioned in section \ref{sec:ARPES} are a possible explanation for the anomalous dispersion of the topmost band.
Fig. \ref{fig:AltreEnergie}, which reports the ARPES intensity plots and their second derivative for photon energies of 60 and 75 eV, shows that no new band at E$_F$ shows up at other photon energies.

Missing adatoms in the center region of the unit cell are the dominant defects found on our well ordered Si(111)--7$\times$7 surfaces. The effect of such vacancies on the electronic structure is shown in Fig. 19 which reports the map of the differential conductance between -0.1 V and -0.2 V on the left, and the topographic image of the same area containing a vacancy on the right. The dI/dV signal - approximately proportional to the integrated LDOS between -0.1 and -0.2 eV- is normalized to its maximum value. The signal on the FCo adatoms that are first nearest neighbors of the vacancy (red circles indicated by the left and right arrows) is about half of the signal on the FCo adatoms far from the defect (other red circles).  Moreover the signal on the vacancy is substantially less than that of the relative adatom.  Therefore a single vacancy substantially decreases the LDOS near E$_F$ on at least three sites. Coherently with this finding the photoemissions spectra show that the intensity of the peak at -0.12 eV is the most dependent on the sample preparation and that it reaches the highest intensity only on the cleanest samples after a proper annealing at 1100 K. The ARPES intensity plot in Fig. 19(c), measured in a sample with defect density so high that the -0.12 eV peak was attenuated by a factor two, shows that the gap and the band dispersion are not strongly affected by the defect density in comparison with Fig. 7 (d). The noise in the plot is caused by the short acquisition time used.

\bibliography{apssamp}

\begin{thebibliography}{72}
\bibitem [{not({\natexlab{c}})}]{notaEF}%
  \BibitemOpen
  \href@noop {} {} ({\natexlab{c}}),\ \bibinfo {note} {the value of E'$_F$ is
  referred to the best estimate of E$_F$}\BibitemShut {NoStop}%
\end{thebibliography}

\end{document}